\documentclass[preprint,12pt]{elsarticle}

\usepackage{amssymb}
\usepackage{float}
\usepackage{graphics}
\usepackage{subcaption}
\usepackage{gensymb}
\usepackage{enumitem}
\usepackage{xcolor}
\usepackage{rotating}

\journal{Computers in Industry}

\begin{document}

\begin{frontmatter}



\title{A Semantic-driven Approach for Maintenance Digitalization in the Pharmaceutical Industry}

\author[inst1,inst2]{Ju Wu}
\affiliation[inst1]{organization={ICT for Sustainable Manufacturing Group, EPFL},%Department and Organization 
            city={Lausanne},
            postcode={1015}, 
            country={Switzerland}}
\author[inst1]{Xiaochen Zheng\corref{cor1}}
\ead{xiaochen.zheng@epfl.ch}
\cortext[cor1]{Corresponding author.}
\author[inst2]{Marco Madlena}
\author[inst1]{Dimitrios Kyritsis}
\affiliation[inst2]{organization={Engineering \& Maintenance Department, Merck Serono},%Department and Organization
            city={Vevey},
            postcode={1809}, 
            country={Switzerland}}
            
\begin{abstract}
The digital transformation of pharmaceutical industry is a challenging task due to the high complexity of involved elements and the strict regulatory compliance. Maintenance activities in the pharmaceutical industry play an essential role in ensuring product quality and integral functioning of equipment and premises.
This paper first identifies the key challenges of digitalization in pharmaceutical industry and creates the corresponding problem space for key involved elements.
A literature review is conducted to investigate the mainstream maintenance strategies, digitalization models, tools and official guidance from authorities in pharmaceutical industry. Based on the review result, a semantic-driven digitalization framework is proposed aiming to improve the digital continuity and cohesion of digital resources and technologies for maintenance activities in the pharmaceutical industry. A case study is conducted to verify the feasibility of the proposed framework based on the water sampling activities in Merck Serono facility in Switzerland. A tool-chain is presented to enable the functional modules of the framework.
Some of the key functional modules within the framework are implemented and have demonstrated satisfactory performance. As one of the outcomes, a digital sampling assistant with web-based services is created to support the automated workflow of water sampling activities. 
%in addressing real challenges derived from user requirements and continuous feedback from water sampling activities, which provides evidence of the effectiveness of the proposed framework and its potential for ongoing development and enhancement. 
The implementation result proves the potential of the proposed framework  to solve the identified problems of maintenance digitalization in the pharmaceutical industry. 
%It is generic and agile which can be easily adapted to specific enterprise use cases.
\end{abstract}


% \begin{highlights}
% \end{highlights}

\begin{keyword}
Semantic technology \sep Pharma 4.0 \sep Digitalization \sep Pharmaceutical industry \sep Maintenance \sep Workflow automation 
% %% MSC codes here, in the form: \MSC code \sep code
% %% or \MSC[2008] code \sep code (2000 is the default)
\end{keyword}

\end{frontmatter}


\section{Introduction}
%The Industry 4.0 revolution is powered by prevalent technologies integrates multiple IT-driven concepts. The ``Smart Factory", equipped with sensors, actors, and autonomous systems, builds the holistic digital models of products and factories. The ``Cyber-physical System" merges physical and the digital spaces. For instance, the process parameters of the physical object like the mechanical components underlying a physical wear and tear are recorded digitally for the purpose of the preventive maintenance. The change towards the decentralized ``Self-organization" comes along with a decomposition of classic production hierarchy in the increasingly-decentralized manufacturing systems. The individualized and connected processes of distribution and procurement are handled by a variety of channels. The new systems to match up with the individualized development of products and services will be constructed by the approaches of open innovation and product intelligence. Essentially, the design of new industrial manufacturing processes ought to meet the social responsibility such as sustainability and resource-efficiency, which are regarded as the fundamental framework factors. 
In the era of the Industry 4.0, new challenges concerning the discipline of business and information systems engineering (BISE) appear with regards to integration, automation and decentralization of enterprise information systems in the existing fields of application such as innovative reference architecture modelling, Business Intelligence (BI) and Enterprise Resource Planning Systems (ERP) approaches and the digital product memories of the full lifecycle \cite{lasi_fettke_kemper_feld_hoffmann_2014}.

The pharmaceutical sector is well-known for its characteristics ranging from high risk, rigorous rules, long and investment-intensive R\&D periods to high-profit margins and significant marketing \cite{scherer2000pharmaceutical,sabouhi2018resilient}. The low-quality pharmaceutical products can be ineffective and even disastrous for all walks of the society, not only the patients. It is necessary to establish a standard framework of regulations and guidelines to mandate all sections of pharmaceutical industry to take proactive actions to guarantee safety and effectiveness of their goods.
As a fusion of Industry 4.0 and quality management in the pharmaceutical industry, the concept of ``Pharma 4.0" has been introduced by International Society for Pharmaceutical Engineering to enhance the exchange of information from the physical process with regulatory authorities by the digitalization in operation models. For the purpose of process optimization, plant performance monitoring, and regulatory compliance within the context of Pharma 4.0, the novel approaches of process analytical technology (PAT), continuous manufacturing and digitalization of pharmaceutical process need to advance \cite{sharifzadeh_2022}. 
%An intelligent automation platform dependent on the middleware between the field devices and enterprise databases is proposed in \cite{coito2020middleware} for the real-time scenarios. A prototype in pharmaceutical industry enhances the platform with the dynamic scheduling algorithms and manifests the interoperability and real-time reaction to changes. The authors of \cite{fertier2021managing} introduces a risk-managing and data-driven architecture i.e., AIC (Acquire, Interpret, Contextualize) information system to improve the situation awareness and resilience within a French pharmaceutical supply chain. Record-keeping for the products and their attributes across the whole production process and supply chain is essential for quality control. A traceability system based on the semantic model is designed and validated in \cite{alonso2016towards} by monitoring the pivotal spatial and temporal points in the pharmaceutical context.

Maintenance plays an indispensable role in the pharmaceutical industry and involves a collection of inter-related processes, operations, organisations and collaborations of various stakeholders. Typical maintenance activities include scheduled sampling of raw materials, intermediate products, and final products to ensure product quality control and assurance. Maintenance management during a system’s full lifecycle invokes the technical, administrative, and managerial actions to ensure the functional utility of the on-site plants. Maintenance activities in the pharmaceutical industry encompass the upkeep of equipment and manufacturing assets to ensure their optimal performance and reliability throughout their lifecycle. These activities play a critical role in maintaining the quality of pharmaceutical products, ensuring compliance with Good Manufacturing Practices (GMP), and minimizing downtime and operational risks. Maintenance activities for the lifecycle management of assets can be classified as reactive or proactive in terms of the way to handle the risks that have potential to manifest as failure. Once a failure induced by the latent risk has happened, reactive maintenance takes reactive measures to correct the emergency issues and malfunctions. Proactive maintenance strategies pay more attention to detect the risks that may lead to the failure and degradation of the functionality of the system by taking calendar-based calibration and inspection of the assets, measuring and recording prognostic information of the components and production processes, to identify potential equipment failures and schedule maintenance interventions. Laboratory Information Management System (LIMS) automates analytical laboratory processes and operations by integrating the management and control of sample and data storage, standards, test results, reporting, laboratory personnel, instruments, and workflow automation \cite{skobelev2011laboratory}.
%\subsection{Good manufacturing practice within pharmaceuticals}\label{sec:1.2}

Good manufacturing practice (GMP) is a globally recognized system for the regulation and management of pharmaceutical product manufacturing and quality control testing that provides proper design, monitoring, and control over manufacturing processes and facilities \cite{world2003expert}. These quality-oriented regulations are implemented to reduce or eliminate errors, impurities and contamination existing in the current pharmaceutical manufacturing and business systems by the relevant regulatory authorities in each country such as the Food and Drug Administration (FDA) in the US, the European Medicines Agency (EMA) and the Medicines and Healthcare Products Regulatory Agency (MHRA) in the UK. The quality-based operations controls such as the record-keeping, staff qualifications, sanitation, equipment verification and cleanliness, and process validation are all covered under GMP rules \cite{patel2008pharmaceutical}.
The Five Principles (5Ps) of GMP in the pharmaceutical industry can be decomposed as five key elements, including People, Process, Procedures, Premises and Equipment, and Products. Each of them has corresponding digitalization requirements.
Considering the context of Pharma 4.0 and the requirements of the key elements, this study aims to address the following digitalization challenges existing in the maintenance activities of the pharmaceutical industry:
\begin{itemize}
    \item  \textbf{Knowledge management} covers explicit (documented), implicit (human skills and capabilities), and computerized \cite{nonaka2009knowledge} knowledge in maintenance. Extracting knowledge is challenging and there is a risk of losing information generated from personalized experiences.  
    
    \item \textbf{Data interoperability} requires arranging complex and partially competing standards on multiple levels covering legal, organizational, semantic, and technical \cite{burns2019review,eif_brochure}. It presents challenges in dealing with data heterogeneity, flexibility, and complexity \cite{moalla2018data}. 
    
    \item \textbf{Workflow automation} uses procedural rules to improve decentralized decision-making, efficiency, and monitoring \cite{stohr2001workflow}. Incomplete digitalization and complex stakeholder cooperation limit automation of maintenance workflows.  
    
    \item \textbf{Training programs} are essential in pharmaceutical industry. Maintenance staff must be trained on emergency operations, compliance, and safety rules. Traditional training systems lack content update and employee feedback. More efficient interactive e-learning platforms are required during digitalization.     
\end{itemize}

\begin{sidewaystable}
    \centering
%\begin{table}[H]
\hspace{-3.25cm}
\resizebox{20cm}{!}{%
\begin{tabular}{|l|l|l|l|l|}
\hline
 &
  Knowledge management &
  Data interoperability &
  Workflow automation &
  Training programs \\ \hline
People &
  \begin{tabular}[c]{@{}l@{}}Extract tacit knowledge like \\ know-how from the experts \\ and transfer them to the \\ newcomers\end{tabular} &
  \begin{tabular}[c]{@{}l@{}}Clear responsibilities matrix, \\ efficient data exchange and \\ communication structure\end{tabular} &
  \begin{tabular}[c]{@{}l@{}}Free employees from \\ manual and repetitive work\end{tabular} &
  \begin{tabular}[c]{@{}l@{}}Customized training based \\ on different roles, integrate \\ subjective experience and \\ feedback\end{tabular} \\ \hline
Process &
  \begin{tabular}[c]{@{}l@{}}Structured and segmented \\ process documentation and \\ complete deviation records \\ including who, when, where \\ and how\end{tabular} &
  \begin{tabular}[c]{@{}l@{}}Unification of elements \\ definition, requirements \\ standard corporate policies \\ and legal framework among \\ sectors\end{tabular} &
  \begin{tabular}[c]{@{}l@{}}Organization of \\ interrelated processes into \\ an automated workflow \\ coordinated by a set of \\ procedural rules\end{tabular} &
  \begin{tabular}[c]{@{}l@{}}Cutting-edge IT technologies \\ supported interactive \\ demonstration, content updates \\ due to process changes\end{tabular} \\ \hline
Procedures &
  \begin{tabular}[c]{@{}l@{}}Detailed explanation of \\ design principles for iterative \\ procedure updates and \\ version tracking\end{tabular} &
  \begin{tabular}[c]{@{}l@{}}Common legislative \\ framework and draft \\ principles for procedural \\ blueprints\end{tabular} &
  \begin{tabular}[c]{@{}l@{}}Records of performance \\ improvement goals and \\ auxiliary tools\end{tabular} &
  \begin{tabular}[c]{@{}l@{}}Clear structure with \\ decomposable modules \\ easy to follow step by step\end{tabular} \\ \hline
\begin{tabular}[c]{@{}l@{}}Premises \\ and \\ equipment\end{tabular} &
  \begin{tabular}[c]{@{}l@{}}Semantic descriptions of \\ interconnections with other \\ elements, changes history \\ tracking\end{tabular} &
  \begin{tabular}[c]{@{}l@{}}Share necessary information \\ in a common format with \\ other divisions\end{tabular} &
  \begin{tabular}[c]{@{}l@{}}Take priorities on the \\ safety rules and \\ contamination prevention \\ measures\end{tabular} &
  \begin{tabular}[c]{@{}l@{}}Thoroughly master how to safely \\ operate functional equipment in \\ the field and handle emergencies\end{tabular} \\ \hline
Products &
  \begin{tabular}[c]{@{}l@{}}Semantically link all phases \\ of the product with related \\ elements of all stages\end{tabular} &
  \begin{tabular}[c]{@{}l@{}}Communicate important \\ product information \\ throughout the lifecycle with \\ other departments\end{tabular} &
  \begin{tabular}[c]{@{}l@{}}Weigh performance \\ metrics and comply with \\ obligatory regulations\end{tabular} &
  \begin{tabular}[c]{@{}l@{}}Hazardous products tips and \\ interactive demonstrations with \\ multimedia support\end{tabular} \\ \hline
\end{tabular}%
}
\caption{The resulting problem space}
\label{tab:problem_space_1}
%\end{table}
\end{sidewaystable}

As presented in Table.~\ref{tab:problem_space_1}, a problem space is created to better explain how the maintenance activities in the pharmaceutical industry should conform to the 5Ps of GMP according to the above-mentioned challenges.
%The proposed semantic-driven digital transformation framework will address the above problem space in the rest of our work.  
Aiming at the problems identified in the problem space, this paper proposes a semantic-driven digitalization framework for maintenance strategies in the pharmaceutical industry. The main contributions of this paper are:
\begin{itemize}
    \item Proposing a human-centric digitalization framework dedicated to the maintenance strategies in the pharmaceutical industry covering the five key elements of GMP.
    %and coordinated by the digital modules explicitly modelling semantic relationships on a common and formally structured basis.
    \item Analyzing the key enabling technologies to implement the proposed digitalization framework focusing on the perspectives of GMP key elements and functional components.
    %\item Developing the digital training modules co-evolving with other modules (e.g., Knowledge Management Module, Maintenance Ontology Module) hone up the staff's skills throughout their careers, adapt them to the changing working conditions, and further decreases the operational risks in safety and hygiene.
    \item Designing a tool-chain for the proposed digitalization framework corresponding to its functional modules.  
    \item Verifying the proposed digitalization framework by developing a web-based digital sampling assistant to support maintenance activities in a real case study.
\end{itemize}

%The rest of this paper is structured as follows: Section \ref{sec:LR} first reviews the existing maintenance strategies within pharmaceutical industry, the related digital tools, and pharmaceutical digitalization framework. Section \ref{sec:2} proposes the schematic diagrams and details of semantic-driven digitalization framework of maintenance strategies as well as the enabling technologies in the pharmaceutical industry and discuss its relations with other models. Section \ref{sec:4} uses the routine water sampling activities for case study. Finally Section \ref{sec:5} summarizes the main achievements of the paper and makes the conclusions.

\section{Literature Review}\label{sec:LR}
This section first reviews the relevant maintenance strategies and tools in the pharmaceutical industry. The limitations of existing strategies are then analyzed. The digitalization is reviewed in the end of this section leading to the main technical contributions of this paper.  

\subsection{Maintenance strategies within pharmaceuticals}
\label{sec:1.3}
%There is a lack of universal maintenance strategies within pharmaceutical industry nowadays. In most cases, the initial development and continuous improvement of these strategies are heavily dependent on the improvisational ideas combined with the current working progress and conditions. 
Although a variety of digitalization technologies have been widely used in the maintenance activities for different local tasks, there are few systematic approaches to coordinate these digital resources and tackle with the digitalization challenges, especially those in the problem space shown in Table.~\ref{tab:problem_space_1}. Some systematic maintenance strategies as well as the relevant official guidelines recommended by the authorities are reviewed as follows.

\begin{itemize}
    \item \textbf{Integrated Management System (IMS)} was suggested as an optimal maintenance strategy and managerial decision-making tool to enhance competitiveness, safety, and efficiency, while reducing costs and failures in the pharmaceutical industry. It considers maintenance levels, work planning and scheduling, maintenance management systems, and required staff comprehensively \cite{felice_petrillo_autorino_2014}. As illustrated in Fig.~\ref{ims_flow_1}, the IMS consists of four steps including a preliminary study, maintenance organizational arrangements, maintenance management process and maintenance improvement procedures.
    
    \item \textbf{Condition-based maintenance (CBM)} framework was proposed to improve abnormal condition management, prevent unexpected deviations and production downtime \cite{ganesh2020design}. It utilizes process knowledge and real-time data through the Real Time Process Management (RTPM) and Process Analytical Technology (PAT) to create an integrated system for process monitoring, material tracking, fault and knowledge management, risk assessment of sensor networks, and supervisory control of critical process parameters (CPPs) and quality attributes (CQAs). The data flow diagram of CBM involves three steps: 1) collecting and contextualizing real-time process operating data; 2) qualitatively and quantitatively assessing potential failure modes based on real-time and historical data; and 3) alerting or notifying the responsible operators and supervisors when evidence of an abnormal condition is detected. 
    
    \item \textbf{Quality Risk Management (QRM)} is a systematic process defined by the FDA for assessing, controlling, communicating, and reviewing risks to drug product quality across its lifecycle \cite{guide_risk}. QRM involves interdisciplinary teams composed of experts from different divisions. The process includes identifying risk issues, assessing risks with quantitative or qualitative measures, designing strategies to reduce risks to an acceptable level, and communicating risk information at any stage of the process. The overview of the QRM process is illustrated in Fig.~\ref{risk_mgmt_1}. The decisions can be made based on the gathered information at any points of the flowchart to modify the configuration of risk model or even terminate the process. 
    
\end{itemize}

%\subsubsection{Limitations of current strategies}
Current maintenance strategies struggle to meet the proposed problem space summarized in Table.~\ref{tab:problem_space_1}, their main limitations are summarized as follows:
\begin{itemize}
    \item The maintenance strategy systems described above are lack of universality to cover all the five key elements of GMP, and they lack the scalability to coordinate their digital modules and resources throughout the full lifecycle.
    \item They lack digital modules to explicitly model semantic relationships on a common and formally structured basis. With the evolution of the project, some abbreviations, acronyms and technical terms may cause ambiguity, affecting the communication and cooperation between different departments.
    \item Their lack of employee-focused digital training modules weakens robustness against changing work environments and increases operational risks relevant to safety and hygiene due to insufficient mastery of domain-specific knowledge and experience.
\end{itemize}
Maintenance activities require an innovative semantic-driven digital transformation framework to reshape cooperation between digital modules for better process and task planning, people and resource coordination, failure prevention, data-driven decision-making, and more.

\subsection{Digitalization within pharmaceuticals}
\begin{figure}[H]
    \centering
  \includegraphics[width = 0.9\textwidth]{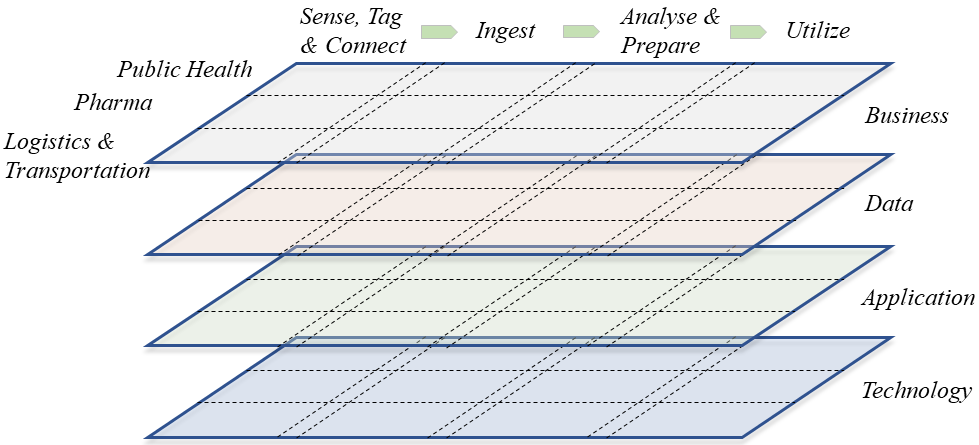}
  \caption{Fundamental structure of the pharma-specific digitalization reference architecture \cite{chircu2017reference}.}
  \label{DT_RA_o1}
 \end{figure}

%Digital transformation reveals new technological solutions and approaches to solving the problems in activities of pharmaceutical enterprises through evolving digitalization technologies, e.g., artificial intelligence (AI), robotic process automation (RPA), IoT, surveillance, social computing, cloud computing, cognitive computing (CC), natural language processing (NLP), data mining, sentiment analysis, human–computer interaction, image processing, geographic information systems, AR, VR, 5G network evolution, bio-metrics, electronic data interchange, etc \cite{gbadegeshin2019effect}. 
The main drivers for the adoption of digital transformation in the pharmaceutical industry can be summarized as reducing costs, improving performance indicators and internal efficiency, promoting smart manufacturing processes, enhancing compliance of policies and standards, increasing work satisfaction for employees \cite{gbadegeshin2019effect}. 
A pharma-specific reference architecture for digitalization, as illustrated in Fig.~\ref{DT_RA_o1}, was developed in \cite{chircu2017reference}. It covers four TOGAF (The Open Group Architecture Framework) standard domains (technology, application, data and business architecture), four digitalization steps (Sense, Tag and Connect; Ingest; Analyze and Prepare; Utilize) and three industry perspectives (logistics and transportation; pharma; public health). 
%The architecture implications of the three emerging digital transformation technologies, i.e., IoT, CC and AR are discussed. IoT assigns people and objects such as sensors, and actuators unique identifiers to enable their interconnection and feedback of their status and surrounding environment; CC combines intelligent algorithms and evolving knowledge to process massive volumes of structured and unstructured data, enabling data analysis and real-time decision making; AR systems increase user perception and interaction with the adjacent real world by combining and aligning real and virtual objects in the real environment in real time \cite{chircu2017reference}.

Regarding the four successive digitization process stages in Fig.~\ref{DT_RA_o1}, the first stage (Tag, Sense \& Connect) consists of tagging non-living or living things with unique IDs and using sensors to identify, locate, track and measure their geospatial and postural attributes and surrounding environments properties. It also includes connecting tagged things over a wireless or wired network to allow them to locate and identify and interact with each other. The second stage (Ingest) involves transferring all data captured in the first stage to a central repository for data storage. The third stage (Analyse \& Prepare) includes analytical data processing operations. The last stage (Utilize) involves managing existing objects or processes, developing new products and services based on the first three stages \cite{chircu2017reference}. In practice, separate reference architectures are built to accommodate the complexity and evolution of each technology.

This reference architecture specifies the key enabling technologies for digital transformation in pharmaceutical industry. However, despite the strengths of the pharma-specific reference architecture, it does not address all the key challenges identified in the proposed problem space for maintenance activities in the pharmaceutical industry. These unsolved challenges include: 
\begin{itemize}
    \item Inadequate knowledge management: The reference architecture does not provide a comprehensive solution for extracting and managing knowledge in maintenance activities. There is a lack of emphasis on capturing explicit and implicit knowledge, such as know-how from experts and personalized experiences. Failure to address this challenge may result in the loss of valuable information and hinder the transfer of expertise to newcomers.
    \item Limited data interoperability: Although the reference architecture acknowledges the importance of data interoperability, it falls short in effectively addressing the complexities and competing standards in the pharmaceutical industry. Dealing with data heterogeneity, flexibility, and complexity remains a challenge, hindering seamless data exchange and communication across different departments and stakeholders.
    \item Insufficient workflow automation: While the reference architecture recognizes the significance of workflow automation, it does not provide a comprehensive solution for coordinating and automating maintenance workflows. The current digitalization efforts in maintenance activities often lack integration and fail to address complex stakeholder cooperation, limiting the potential for efficient and automated workflow management.
    \item Ineffective training programs: The reference architecture does not adequately address the need for more efficient and interactive training programs in the pharmaceutical industry. Traditional training systems lack content updates and employee feedback mechanisms, resulting in limited adaptability to changing working conditions and potential gaps in domain-specific knowledge and experience.
\end{itemize}

%This study adopts semantic technology and combines it with this reference architecture to propose a more generic and high-level digitalization framework considering the entire lifecycle of the system.
%Although \cite{chircu2017reference} has constructed the technology-specific reference architectures, there is still lack of the generic and high-level digitalization frameworks within pharmaceuticals, especially for the maintenance activities.

\section{Semantic-driven Digitalization Framework of Maintenance Strategies}\label{sec:2}

A semantic-driven digitalization framework for maintenance strategies in pharmaceutical industry is proposed to tackle with the problem space summarized in Table.~\ref{tab:problem_space_1}.   

\subsection{The proposed framework and its components}

\begin{figure}[H]
    \centering
  \includegraphics[width = 0.9\textwidth]{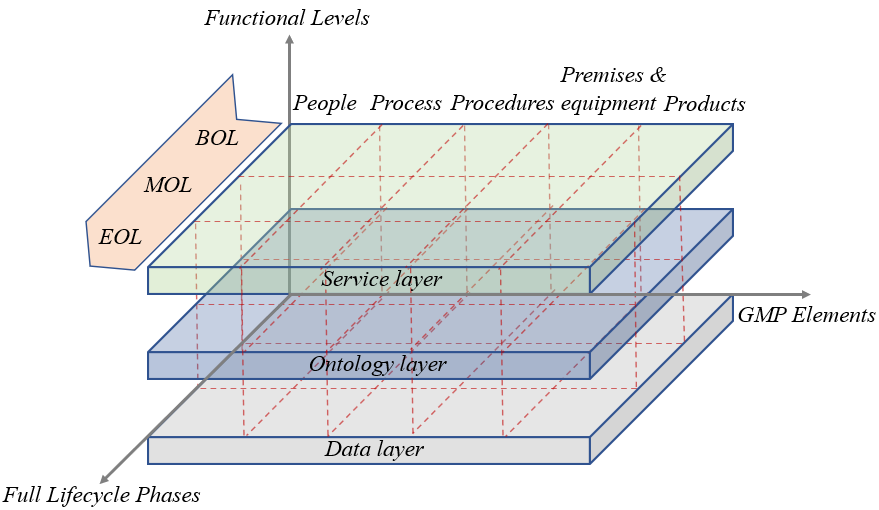}
  \caption{General architecture of the semantic-driven digitalization framework for maintenance strategies in pharmaceutical industry.}
  \label{DT_RA_1}
 \end{figure}

The proposed semantic-driven digitalization framework for maintenance strategies in pharmaceutical industry is illustrated in Fig.~\ref{DT_RA_1}. It is inspired by the Cognitive Digital Twin (CDT) \cite{zheng2021emergence} conceptual architecture and the three-layer (i.e., data-ontology-service) Manufacturing Model (MfM) methodology proposed by \cite{mas2018preliminary}. The architecture of the framework refers to RAMI 4.0 \cite{hankel2015reference}. The proposed framework is composed of three dimensions: 1) the five pivotal elements of GMP (i.e., people, process, procedures, premises \& equipment, products); 2) three functional layers (i.e., service layer, ontology layer, and data layer); and 3) full lifecycle phases including beginning-of-life (BOL, e.g. problem identification and analysis, solution feasibility study, prototype design and testing, resources scheduling), middle-of-life (MOL, e.g. operating, periodic review and inspection, regular feedback, revision and improvement, archiving of data and information) and end-of-life (EOL, e.g. programs cancellation, manpower and material resources withdrawal, recycling, re-configuring maintenance activities).

In the pharmaceutical industry, the five GMP key elements are organized into clusters based on various standards, criteria, and metrics. Throughout the lifecycle, a dynamic network of these instances from the five elements collaborates, providing complementary functions and operating in different roles. While the digitalization versions of the GMP elements vary across lifecycle phases, they remain interdependent and evolve synergistically. At the BOL, stakeholders' requirements are analyzed, preliminary solutions are identified, and digital modules are prototyped and tested. The necessary databases, repositories, and software are deployed, and initial services modules are constructed. The integrated digital modules are then evaluated as a whole system, followed by organizing human resources, providing training, and arranging required materials and equipment. At the MOL, decision-makers review and audit the results of digital modules, assessing their operating conditions based on historical data and user feedback. Coordination between modules is revised and improved, while data management platforms are updated. Application-level ontologies and knowledge bases may be adapted, and the functionalities of service layer modules are amended according to new goals. Personnel receive revised training, and material, equipment, and premises are rearranged efficiently yet robustly. At the EOL, performance evaluation reports and user requirements are analyzed to determine whether certain components of the digital framework should be canceled, resulting in the decomposition and recycling of resources for the construction of other digital modules.

\subsubsection{The perspective on GMP elements and functional levels}

\begin{figure}[H]
     \hspace{-9.6mm}
  \includegraphics[width =  1.06\textwidth]{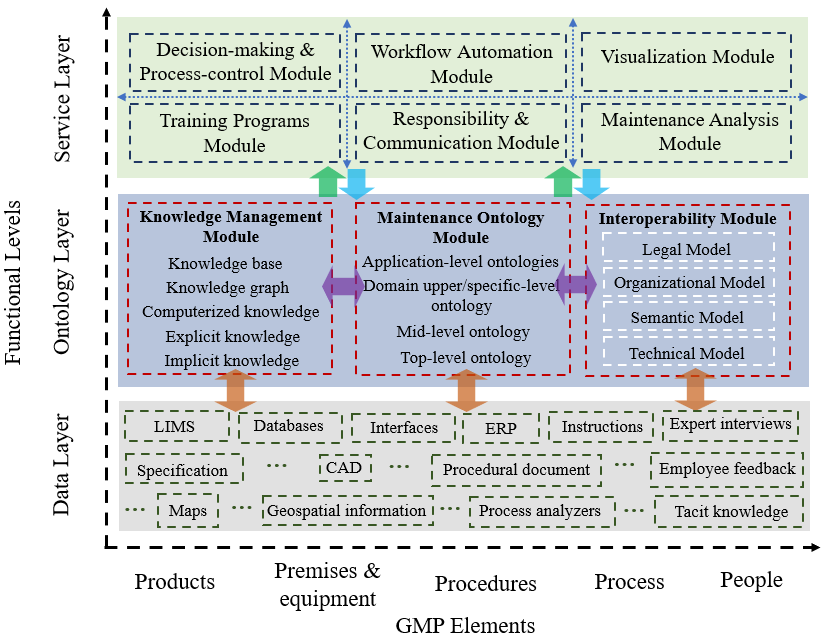}
  \caption{The GMP elements and functional levels plane of the proposed framework.}
  \label{SDT_YZ_1}
 \end{figure}

This perspective of the proposed framework illustrates how the interrelated and self-evolving digitalization modules categorized into different functional levels (i.e., data layer, ontology layer, and service layer) orchestrate the five GMP elements, as shown in Fig.~\ref{SDT_YZ_1}. 

\begin{itemize}
    \item \textbf{The data layer} maintains relevant resources, channels, legacy platforms and software regarding the five GMP elements, e.g. structured and unstructured databases, procedural documents, specifications, instructions, LIMS, ERP, CAD, process analyzers and their associated interfaces, maps and their associated geospatial information, expertise from expert interviews, regular and ad-hoc feedback from on-site employees, tacit knowledge from senior employees, etc, through which valuable data, information and knowledge can be provided and extracted to the ontology layer. 
    
    \item \textbf{The ontology layer} consists of the following three modules: 
    \begin{itemize}       
        \item The knowledge management module uses formal semantics to manage the knowledge base. It organizes computerized, explicit and implicit knowledge from multi-source data extraction and synthesis. The resulting knowledge graph is aligned with ontologies to support knowledge integration, querying, and reasoning. 
        
        \item The maintenance ontology module is based on customized application ontologies for maintenance activities, which are developed according to specific ontology construction principles and conventions. The hierarchical IOF (Industrial Ontologies Foundry) technical principle is adopted in this study which suggests using top-level, mid-level and domain-level ontologies as basis to construct application ontologies. By this way, it can facilitate ontology merging, reusing, inter-operating, collaborative editing and extending.        
        The application-level ontologies are dedicated to certain application scenarios for maintenance activities. Some templates can be defined for further instantiating such as stakeholders communication structure templates, acronyms and terminology unification template, user requirement description template, templates for legal framework, corporate policies and standards, maintenance process template, templates for decision-making strategies, among others.        
        The structural evolution and refactoring of the maintenance ontology module is also inspired by the content of the knowledge management module. 
        
        \item The interoperability module, provides design requirements and guiding principles for the construction of digitalization models of the proposed framework to guarantee data interoperability \cite{eif_brochure}. Some of its components are instantiated based on the maintenance ontology module and knowledge management module. It is composed of the following four models:   
        \begin{itemize}
            \item Legal model: the initial step is to list all relevant legislative provisions, policies, standards, and regulations pertaining to the five pivotal elements of GMP. Performative knowledge and application-level ontologies are built based on this list, including ontologies for legal framework, policies and standards, and practical legal knowledge. Interoperability checks are then conducted to identify any conflicts between these models and the knowledge management module. Limitations and reconciling patches are monitored and updated throughout the lifecycle of the five pivotal elements of GMP.        
            \item Organizational model: clearly-defined and self-evolving relationship networks must be established for stakeholders regarding responsibilities, communication, service and material transfer. These networks are aligned with the five pivotal elements of GMP. All forms of knowledge are documented and specific application ontologies are instantiated from templates of the maintenance ontology module, including responsibility assignment, stakeholders communication, user requirement description, risk management, and maintenance process integration and workflow automation models.        
            \item Semantic model: the raw data, information, and knowledge gathered from the data layer have highly diverse sources and formats. There should be consensus on a description language with publicly accepted conventions, syntax, and format, and the transformation of data in different formats from legacy systems into a common format for better stakeholder communication, information transfer, and data exchange. Then all forms of corresponding knowledge are documented and specific application-level ontologies are instantiated by the templates in the maintenance ontology module.        
            
            \item Technical model: the domain-specific digital applications and legacy information systems are usually developed and maintained by different performative groups in the maintenance department. 
            %These digital tools can neither easily exchange data and information nor synthesize into an centralized interface to increase service accessibility for the personnel, functional integration and workflow automation. 
            There is an urgent need of stronger centralized digital control capabilities with hierarchical permissions for the decision makers to integrate critical process data for monitoring and visualization, statistical maintenance process analysis, data-driven fault diagnosis, automated notification and risk alerts, preventive maintenance strategies generation and process control etc.           
            To support construction of this model, a variety of sub-domain knowledge from the knowledge management module is required and multiple application-level ontologies are built.
            %based on the following templates: real-time data visualization and monitoring model, model for maintenance process analysis and fault diagnosis, model for decision-making strategies and maintenance process control, etc.
        \end{itemize}    
    \end{itemize}
    Feedback from the ontology layer in turn helps to improve and redesign the way information and data are captured and synthesized from the data layer, such as re-configuring questionnaires, redesigning question and answer forms for expert and field operator interviews, updating data exchange formats and legacy databases and information management software.

    \item \textbf{The service layer} is dedicated to the user requirements for different scenarios, which is developed and continuously improved on top of the ontology layer. It consists of six interrelated and coordinated digital service modules as follows:
    
    \begin{itemize}
        \item Decision-making and process-control module: this module coordinates decision-making and process control for maintenance activities pertaining to the five pivotal elements of GMP, organizing physical, digital, and human resources throughout their lifecycle, including quality risk management, PAT system design, emergency response planning, preventive maintenance, personnel and tasks scheduling, and system development for plant process control.
        
        \item Workflow automation module: this module automates maintenance activities through procedural rules determined by decision makers and controlled by the workflow management system. The module is continuously evaluated, tested, and improved, and includes monitoring of compliance with regulations and standards, as well as management of equipment and premises.
        
        \item Visualization module: this module involves design of user experience (UX)/user interface (UI), graphic user interface (GUI) layout design and beautification for digital applications, layout design of the internal BI reports charts, tables and forms. All of the above visualization services need to be human-centric, collectively decided by all the participating stakeholders, and constantly evolving based on users' feedback and new needs. 
        
        \item Training programs module: this module covers a set of training services for the entire lifecycle of the five key GMP elements, beginning with the initial boarding programs targeted at different executive teams. The training programs consist of re-configurable sessions and a flexible structure to meet the customized needs of employees throughout their careers. 
       % Apparently, this module tailors the diversified and customized training plans for the unit decision-makers, software engineers, digitalization managers, and the frontier operators. 
        The critical and error-prone steps of the maintenance process are supported by the cutting-edge IT technologies such as AR, VR and MR. 
        The module contains evaluation forms to mark the operations that are hard to follow or difficult to meet the safety and contamination rules.
        It also offers mixed assessments to evaluate staff's capability. 
        %For instance, flexible quizzes can be used to test the mastery of knowledge of the maintenance activities; investigation and evaluation of the independent handling of operations by the newcomers after the senior staff's hands-on demonstration, especially for high-risk activities.
        The safety and risk prevention are taken as priorities in this training module. %the employees will thoroughly learn to be accustomed with the geospatial conditions such as the safety exits of the premises where they are working in, to properly use personal protective equipment for self-protection, to operate each functional equipment safely on site, and to prevent and respond to possible accidents and emergency conditions. 
        %This part is located at the head for each course and serves as prerequisite for independent hands-on practice.         
        %This module also organizes the training sessions to allow the staff be familiar with the common guidelines and policies of safety and hygiene in pharmaceutical industry such as PAT framework, ICH Q9 guidance and GMP rules of the region in which the enterprise is located, and be notified of their latest updates. 
        The training programs of this module support version control and evolve according to changes such as the new requirements of maintenance processes, platform back-end statistics and feedback from employees.
        
        \item Responsibility \& communication module: this module clarifies the organizational hierarchies by digitally maintaining the responsibilities and communication networks of the personnel, regarding the service and material transfer of the five pivotal elements of GMP involved in the maintenance activities; at the same time, the regular offline and online discussions are planned and conducted by various executive teams to scheme the phased goals, user requirements collectively; and the performance expectations, quality and risk control, periodic checkpoints are reviewed together as well according to the content of the ontology layer.
        
        \item Maintenance analysis module: this module covers all digital strategies, approaches and tools of qualitative and quantitative analysis in the maintenance activities, using the necessary domain knowledge, e.g., PAT tools that establish the multi-factorial relationships between processes and products throughout their lifecycles, risk-mitigation ways in the integrated PAT system, risk handling approaches in the quality risk management system, sensor-based process measurements and configuration of their associated interfaces; historical data collection and statistical algorithms for performance evaluation and failure prediction and detection, legislation and regulation analysis to ensure consistent compliance with policies and standards, personnel performance monitoring and assessment, and more.
    \end{itemize}    
    Correspondingly, the frequent interaction between users and the digital modules of the service layer helps to introduce novel improvement ideas outside the enterprise and stimulate the consistent evolution of the ontology layer.
\end{itemize}

\subsection{Relationship with other models}
\subsubsection{Relationship with RAMI 4.0}
The proposed digitalization framework takes RAMI 4.0 in \cite{hankel2015reference} as guideline to build up the three pillars. The full lifecycle phases of the proposed digitalization framework serves the equivalent roles as the \textit{lifecycle Value Stream (IEC 62890)} dimension of the RAMI 4.0. The \textit{Hierarchy Levels (IEC 62264//IEC 61512)} axis of RAMI 4.0 is replaced by the key GMP elements that denote the different functionalities within factories or facilities regarding the maintenance activities. The \textit{Hierarchy Levels (IEC 62264//IEC 61512)} axis of RAMI 4.0 mainly represents the generic enterprise IT and control systems of Industry 4.0 environment, including work pieces like products and devices, the connection to the IoT and services, the GMP elements of the proposed digital framework detail the people, products, processes, procedures, equipment and premises specifically involved in maintenance activities in the pharmaceutical industry. 
%The \textit{Layers} axis of RAMI 4.0 is replaced by the functional levels in the proposed framework, and they have different divisions for the virtual entities in the digital space. 

The \textit{Layers} axis of RAMI 4.0 decomposes the virtual mapping of a machine into six layers to meet the needs and challenges around the products from a business perspective: the \textit{Business Layer} is located on top of this pillar to take care of the customer wishes and the matching organisation and business processes, the \textit{Functions Layer} details the functions of the products and relevant assets, the \textit{Information Layer} records the necessary information concerning the products, the \textit{Communication Layer} is responsible for the infrastructures and approaches to access the products information and data, the \textit{Asset Layer} clarifies how the products transform and move through the processes in the physical world, the \textit{Integration Layer} concerns the integration of products and assets into the information world by digitalizing them. In contrast, the functional levels of the proposed framework address the digitalization challenges, not limited to the business aspects, which the maintenance activities in pharmaceutical industry are facing, in particular knowledge management, data interoperability, workflow automation and training programs illustrated in Table.~\ref{tab:problem_space_1}, through synergizing the virtual entities, digital resources and data-driven services related to the five key GMP elements into the cohesively interrelated and inter-dependent digital modules, distributed in the three layers (data layer, ontology layer and service layer). The users with diverse roles, responsibilities and permission levels can interact with different parts of the proposed framework shown in Fig.~\ref{DT_RA_1} at any stages.

\subsubsection{Relationship with TOGAF}
The pharma-specific TOGAF reference architecture \cite{chircu2017reference} has three pillars. It follows a different way to organize the digital assets and implement the digital transformation, its \textit{industry perspectives} axis is too broad to focus on the specific GMP elements of the maintenance activities in pharmaceutical industry, its \textit{standard domains} axis covers technology, data, application and business and serves the similar roles as the functional levels of the proposed framework but lack of emphasis on knowledge management and data interoperability, its \textit{digitalization steps} axis corresponds to the full lifecycle phases of the proposed framework but it dismisses integration of diverse digital systems and technologies. By incorporating digital modules and assets relevant to knowledge management, data interoperability, workflow automation, and training programs, the proposed framework aims to solve the key problems that are not fully addressed by the pharma-specific reference architecture in \cite{chircu2017reference}, and it provides a more detailed and specific approach to digitalization, enabling efficient data integration, knowledge sharing, and service delivery throughout the lifecycle of maintenance activities.

\subsection{Enabling technologies}\label{sec:3}
Apart from the cutting-edge digitalization technologies mentioned in the previous sections, some of the fundamental enabling technologies supporting the proposed digitalization framework are introduced below.

\subsubsection{Ontology engineering and knowledge graph}
\begin{figure}[H]
    \centering
  \includegraphics[width = 0.86\textwidth]{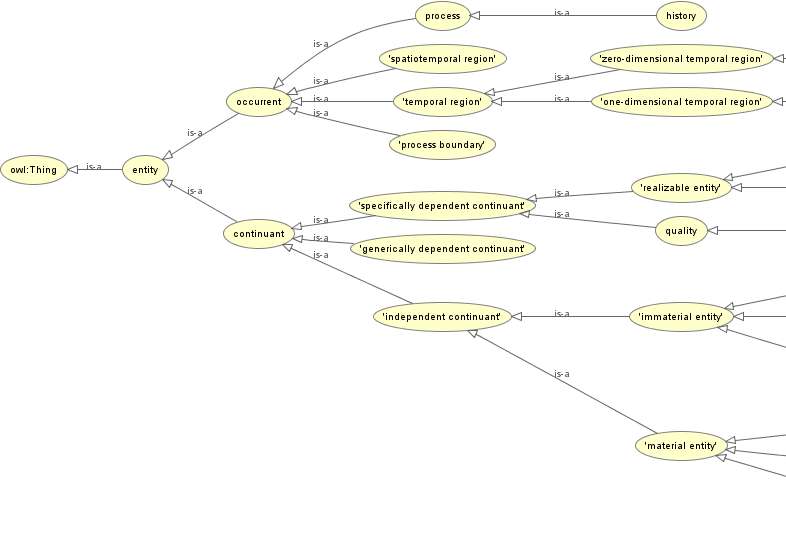}
  \caption{Classes diagram of BFO}
  \label{bfo_graph}
 \end{figure}

Ontology is an efficient semantic modelling tool which can help capture the knowledge of a certain domain with formalized classes and properties. The upper-level ontologies provides groups of formally-defined and properly-structured vocabularies and semantic relations, serving as a common foundation for developing lower-level ontologies such as domain specific ontologies and scenarios-dependent application ontologies \cite{zheng2021emergence}.
Basic Formal Ontology (BFO) is the genuine top-level ontology, which does not contain the terms of specific scientific domains. The ``entity" of BFO denotes anything that exists, has existed and will exist. The ``entity" has two subclasses, i.e., ``continuant" represents the entity that persists, endures and continues to exist through time while maintaining its identity, and ``occurrent" denotes the entity that unfolds itself in time, the start or end of such entity as the boundaries and thresholds, or temporal or spatio-temporal regions in which such entity occurs \cite{arp2015building}. 
Some of the key classes of BFO are visualized in Protégé Ontology Editor as shownFig.~\ref{bfo_graph}.

\begin{figure}[H]
     \hspace{8.6mm}
  \includegraphics[width =   0.86\textwidth]{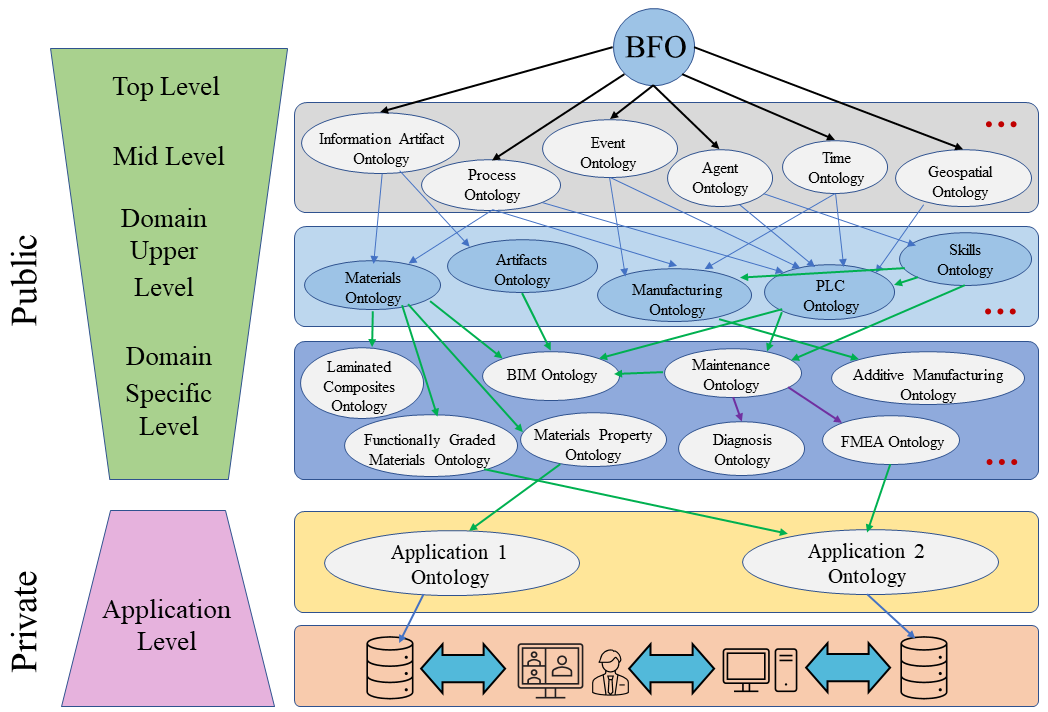}
  \caption{IOF architecture vision. Adapted from \cite{iof_1}}
  \label{iof_arch}
 \end{figure}

 IOF-Core created by IOF, aiming to support data interoperability for industrial manufacturing, co-creates a coherent and consistent development architecture composed of a set of open-sourced ontologies, in which the lower-level ontologies are constructed cumulatively from the upper-level ones in a hierarchical way, as shown in Fig.~\ref{iof_arch}. Derived from the root ontology BFO, IOF-Core includes additional predefined domain upper level ontologies and domain specific level ontologies, which are introduced in \cite{iof_code}.

% The BFO framework has the basic universal-universal relations classified as four types as follows \cite{arp2015building}.

% \begin{itemize}
%     \item foundational relations:
%     \begin{itemize}
%         \item \textit{is\_a}(is a sub-type of)
%         \item \textit{continuant\_part\_of}
%         \item \textit{occurrent\_part\_of}
%     \end{itemize}
%     \item spatial relations:
%     \begin{itemize}
%         \item \textit{located\_in}
%         \item \textit{adjacent\_to}
%     \end{itemize}
%     \item temporal relations:
%     \begin{itemize}
%         \item \textit{adjacent\_to}
%         \item \textit{preceded\_by}
%     \end{itemize}
%     \item participation relations:
%     \begin{itemize}
%         \item \textit{has\_participant}
%     \end{itemize}
% \end{itemize}

%\subsubsection{Knowledge graph}
The developed application ontology can be used as basis to construct a knowledge graph to enable more powerful capabilities for knowledge representation, heterogeneous data integration and reasoning. The knowledge-based system can acquire information from multiple information sources through the ontology and apply a reasoner to derive new knowledge. 
%whose abstract architecture is shown in Fig.~\ref{kg_arch}.

% \begin{figure}[H]
%      \hspace{29.6mm}
%   \includegraphics[width =  0.66\textwidth]{fig14_kg_arch_1.png}
%   \caption{A typical knowledge graph architecture. Adapted from \cite{ehrlinger2016towards}}
%   \label{kg_arch}
%  \end{figure}
 
\subsubsection{Agile developments of digital tools}

The commercial cloud-service based platforms such as Microsoft Power Platform, Google Cloud Platform, Amazon Web Services, featuring low-code and high-flexibility, support agile developments of digital tools and maintenance across the full lifecycle. Their computing resources can be integrated to organize a versatile ecosystem, streamlining the business processes within the enterprise and satisfying the customized organizational needs such as building role/task specific digital tools to handle large amounts of data from different sources and to automate the manual workflow. A typical cloud-based digital tool ecosystem can be constructed which will enable the users to jointly develop digital tools in distributed or centralized way.
\iffalse
\begin{figure}[H]
     \hspace{3.6mm}
  \includegraphics[width =   0.96\textwidth]{fig6_agile_dev_1.png}
  \caption{A typical cloud-services based ecosystems for digital tools development.}
  \label{agile_dev_1}
 \end{figure}
 \fi
\subsubsection{Learning Management System}

A learning management system (LMS) is a software application that automates the administration, tracking, and reporting of training events \cite{ellis2009field}. A robust LMS should centralize and automate administration, use self-service and self-guided services, assemble and deliver learning content rapidly, consolidate training initiatives on a scalable web-based platform, support portability and standards, and enable personalization and knowledge reuse. The LMS will also need to sample data from across the entire IT organization and offer better search engine optimization \cite{davis2009evolution}. The integration between LMS and authoring tools such as Articulate Storyline provides a seamless experience for learners, who can access all their learning content in one place.\\

%Authoring tools such as Articulate Storyline are often used in conjunction with an LMS to create and deliver e-learning content. Content created in authoring tools can be uploaded and managed in an LMS. LMSs may offer built-in authoring tools or integrate with third-party authoring tools. The integration between LMS and authoring tools provides a seamless experience for learners, who can access all their learning content in one place.

\section{Case study} \label{sec:4}

In Merck Serono facility at Vevey (biotech production plant) many samples of purified water and condensed purified steam are collected daily and analyzed  in QC laboratories to certify the quality of the products. The sampling operation takes around 4 hours a day and brings heavy workloads to the Engineering \& Maintenance (E\&M) department.  The target of this case study is to optimize the water sampling process using the proposed digitalization framework and relevant enabling tools. 

It is worth noting that the water sampling activities are also relevant to quality assurance and sometimes are classified as part of quality assurance activities. However, in this case study, water sampling is not purely a quality assurance measure. By regularly testing the quality of the sampled purified water and recording its status, it serves as a crucial means to reflect the performance of the manufacturing systems, including plant assets and equipment. The testing results provide valuable insights into the functioning of the systems and can indicate whether maintenance interventions are required, helping to ensure their continued optimal operation and prevent potential disruptions to the production process. Moreover, in this particular production plant, water sampling is an important task of the maintenance department as a key procedure of the overall maintenance strategy. Therefore, in this study we consider water sampling as maintenance activities.

\subsection{Application scenario}
\begin{figure}[H]
     \hspace{-4.9mm}
  \includegraphics[width = \textwidth]{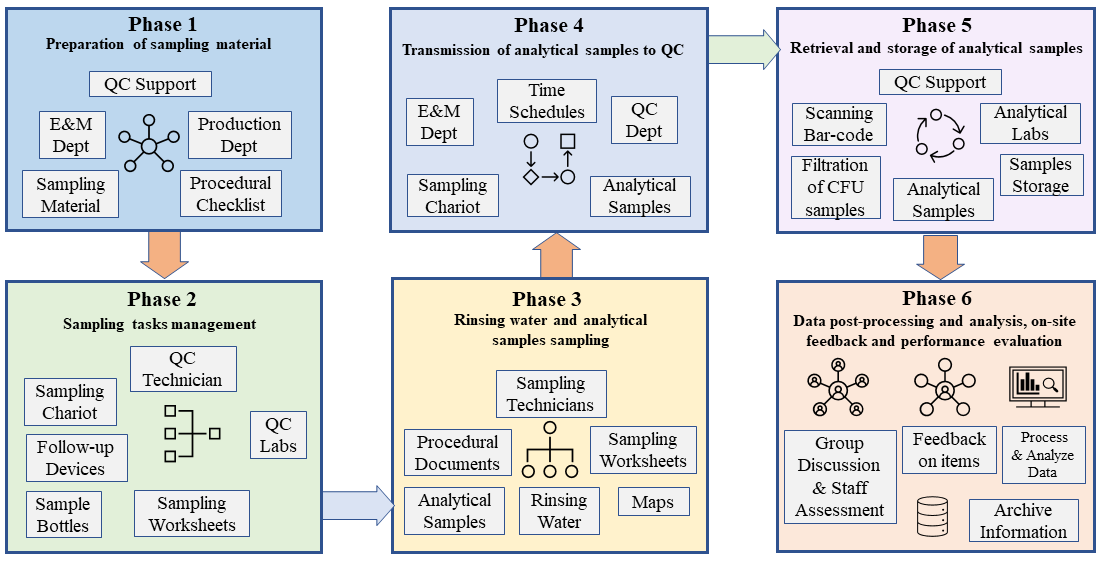}
  \caption{Workflow of water sampling activities and the elements involved.}
  \label{ws_flow_1}
 \end{figure}

\begin{figure}[H]
\centering
\begin{subfigure}{0.45\textwidth}
    \includegraphics[width=\textwidth]{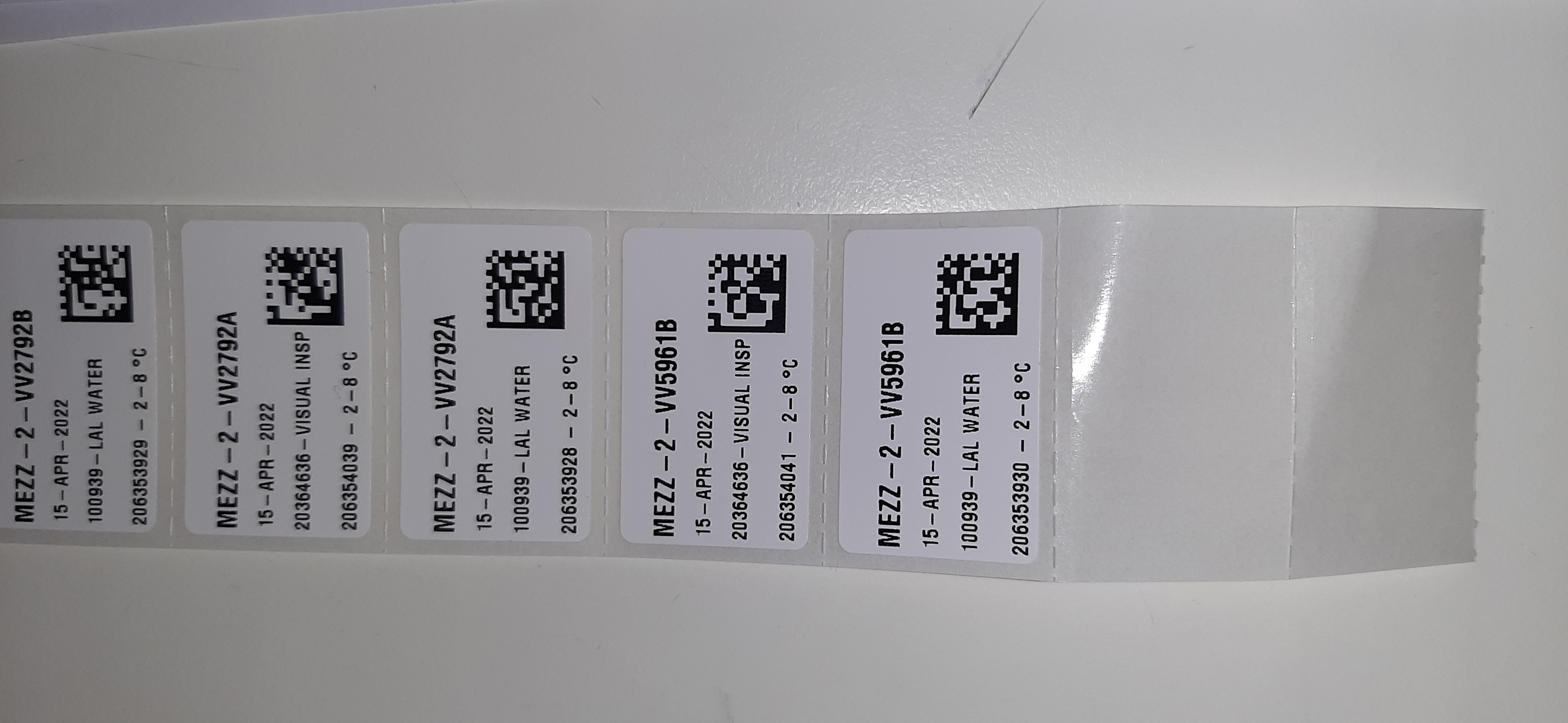}
    \caption{QR code as bar-code in the adhesive sticker for reception of water samples.}
    \label{sampling_figs_label}
\end{subfigure}
\hfill
\begin{subfigure}{0.475\textwidth}
    \includegraphics[width=\textwidth]{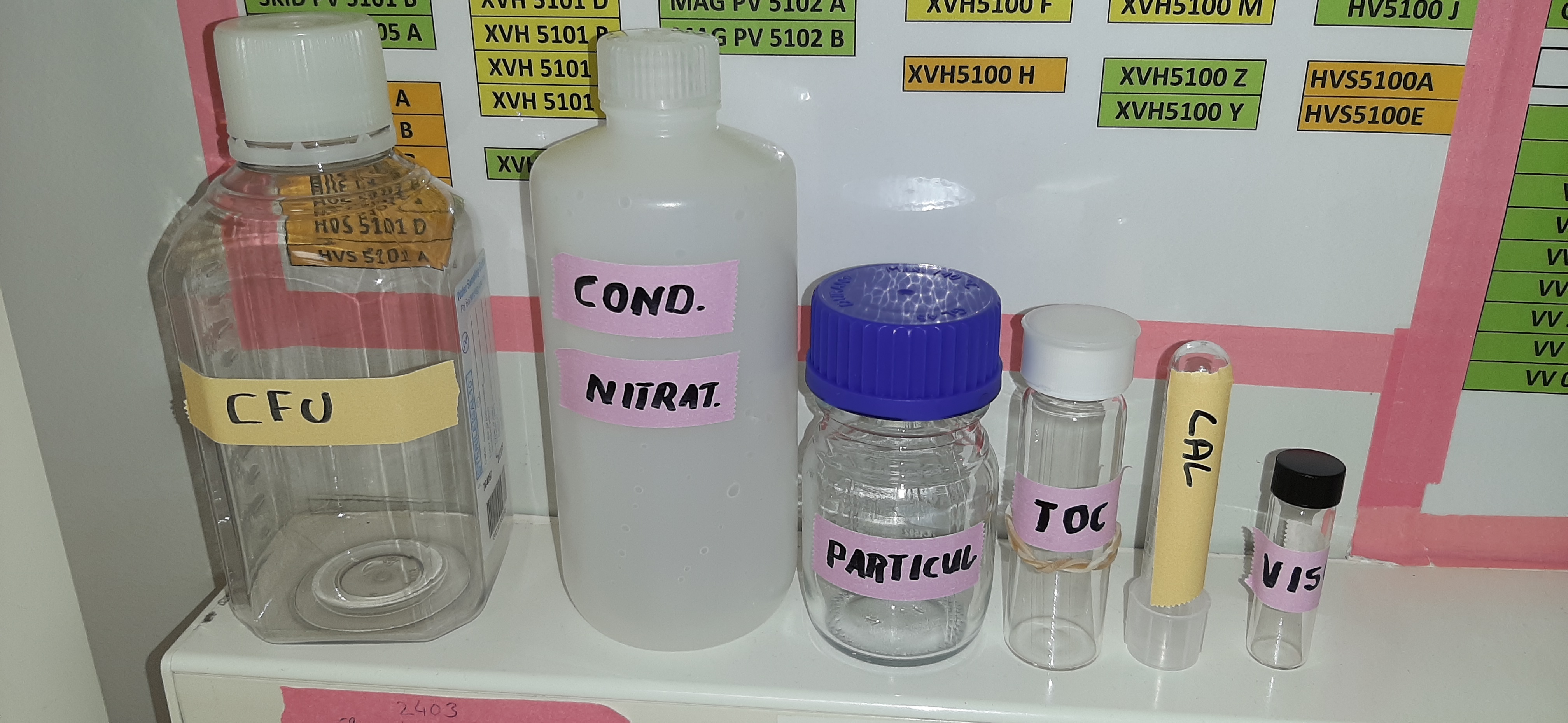}
    \caption{Sample bottles with the colored marks.}
    \label{sampling_figs_bottle}
\end{subfigure}
\hfill
\begin{subfigure}{0.475\textwidth}
    \includegraphics[width=\textwidth]{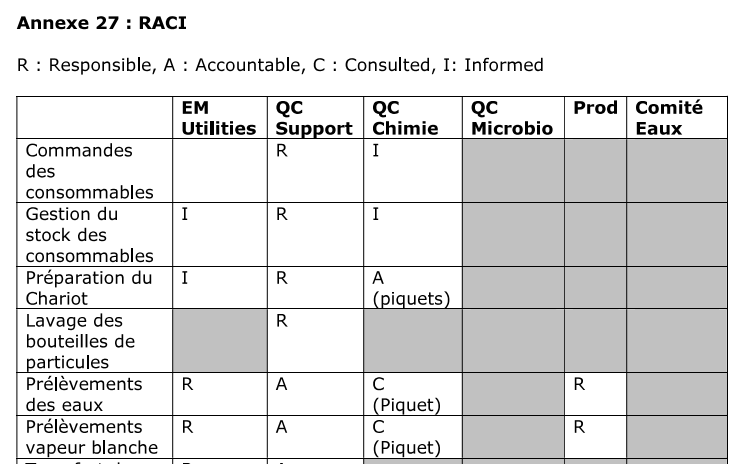}
    \caption{Part of RACI table.}
    \label{sampling_figs_raci}
\end{subfigure}
\hfill
\begin{subfigure}{0.45\textwidth}
    \includegraphics[width=\textwidth]{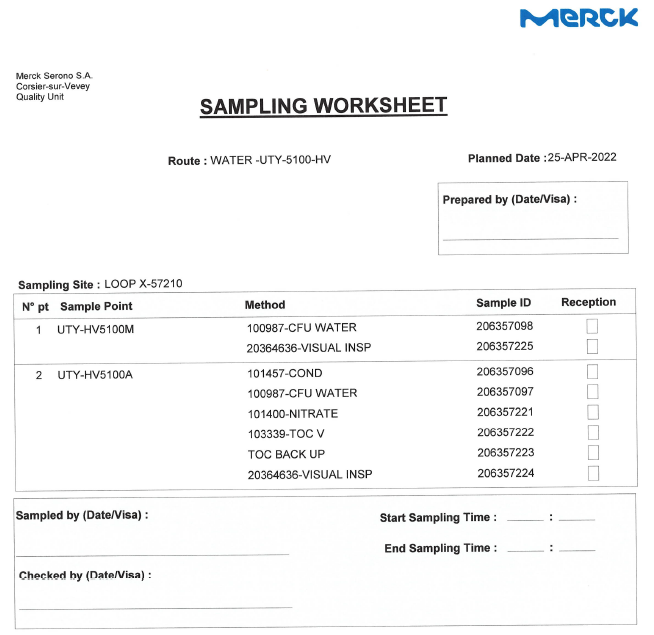}
    \caption{Typical layout of daily printed worksheet.}
    \label{sampling_figs_ws}
\end{subfigure}
\hfill
\begin{subfigure}{0.4\textwidth}
    \includegraphics[width=\textwidth,angle=-90]{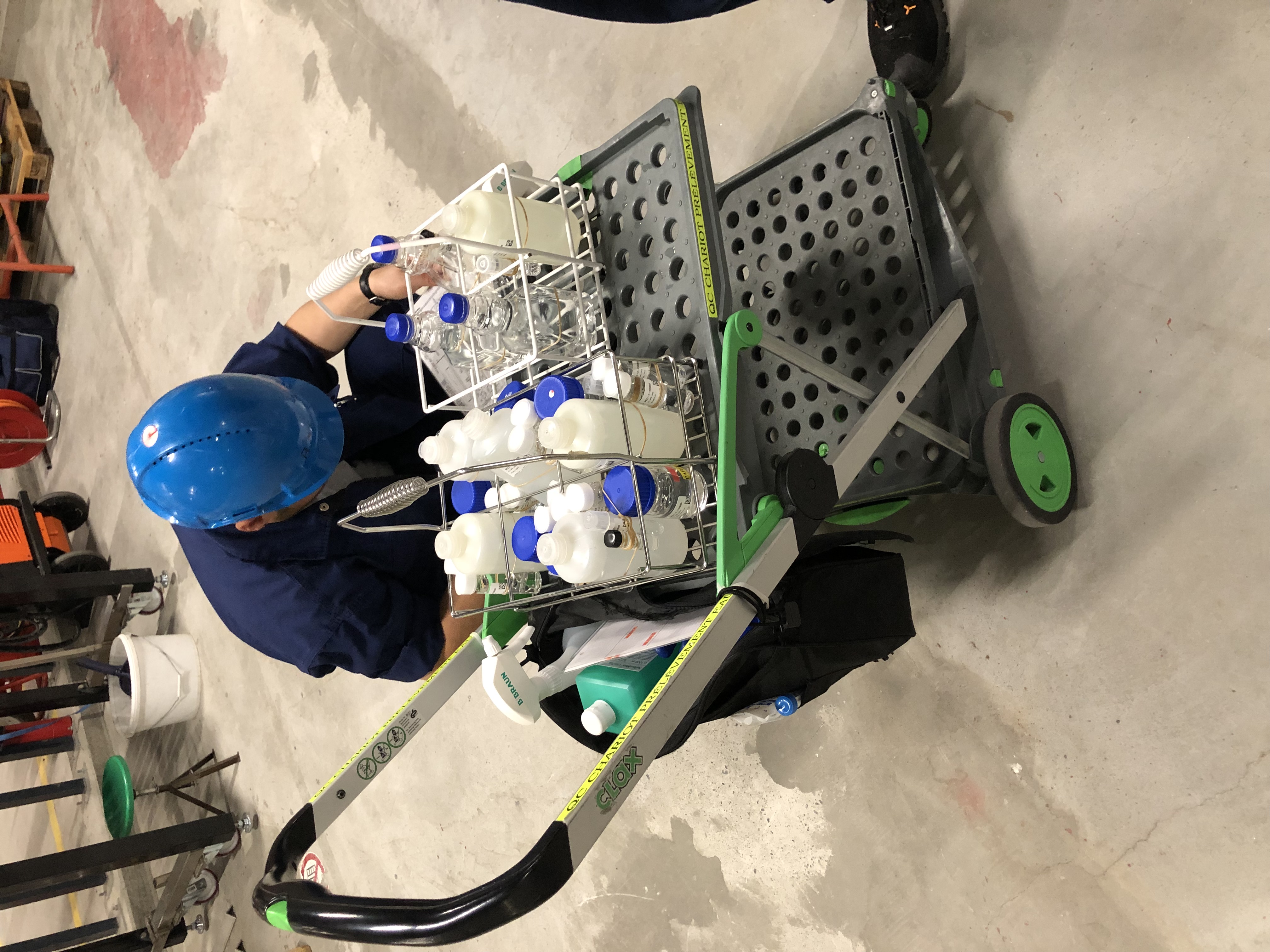}
    \caption{Sampling chariot and technician.}
    \label{sampling_figs_people}
\end{subfigure}
\hfill
\begin{subfigure}{0.475\textwidth}
    \includegraphics[width=\textwidth,angle=-180]{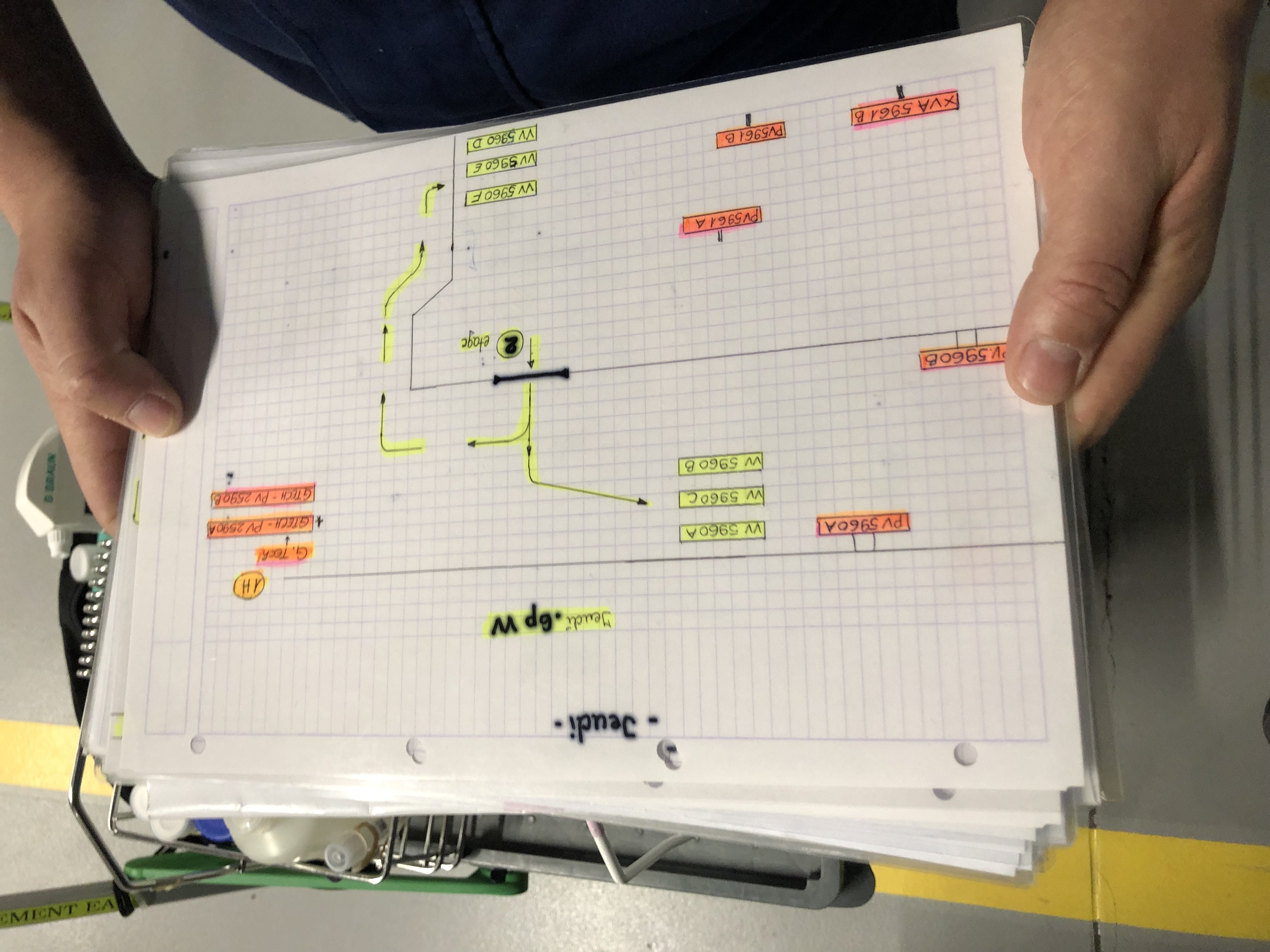}
    \caption{Hand-written maps of sampling point plan for a specific date from the technicians.}
    \label{sampling_figs_map}
\end{subfigure}
\caption{Parts of the images of water sampling activities}
\label{sampling_figs}
\end{figure}

The general workflow of water sampling activities and the elements involved are shown in Fig.~\ref{ws_flow_1} and explained as follows:

\begin{enumerate}[label=Phase \arabic*]
    \item QC Support prepares the sampling material for E\&M, as shown in Fig.~\ref{sampling_figs_people}, according to the detailed activity specification .
    \item QC technician deposits the sample bottles directly on the chariot in QC laboratories with the follow-up devices and sampling worksheets.
    \item The technicians collect analytical samples and the corresponding rinsing water according to the daily sampling worksheet, as shown in Fig.~\ref{sampling_figs_ws}, and the detailed descriptions of the sampling methods. Currently, the technicians rely on the printed or hand-written maps Fig.~\ref{sampling_figs_map} for sampling.
    \item Upon completion of sampling, the E\&M brings back the chariot to QC according to scheduled time. 
    \item QC Support receives the samples by scanning the bar-code, as shown in Fig.~\ref{sampling_figs_label}, and makes the Colony Forming Units (CFUs) available for the filtration. The other samples are distributed to the laboratories for analyses.
   % Water samples can be left for a maximum of 8 hours at room temperature before analysis or stored at +2\degree C/+8\degree C for a longer period depending on different kinds of analytical samples. Note that water samples must imperatively be received by scanning the bar-code shown in Fig.~\ref{sampling_figs_label} and no other method is acceptable.
\end{enumerate}

%As illustrated in Fig.~\ref{sampling_figs_ws}, within the daily printed water sampling worksheet, a sampling batch is defined starting from information i.e., ``prepared by (Date/Visa)" and completing with information i.e., ``sampled by (Date/Visa)", ``checked by (Date/Visa)", ``start sampling time", and ``end sampling time". The rows of N\degree pt" represent the sampling tasks within this sampling batch. The design and layout of sampling batches determines the spatio-temporal division of the overall sampling workloads and mobility of the technicians. The row of sampling tasks ``N\degree pt" can be regarded as the minimum basic task unit, comprised of the column features i.e., ``Sample Point", ``Sampling Dept", ``Sample Description", ``Sampling Execution Date", ``Sampling Zone", ``Sampling Method", ``Sample ID", ``reception" checkbox, etc.

%\subsection{User requirements}
%User requirements are identified through procedure documents analysis and interviews with experts. 
%These user requirements serve as a reference for delineating milestones and brainstorming of all participating stakeholders in the ongoing development of a domain-specific digital framework maintenance strategies for water sampling activities.

\subsubsection{Existing problems}
After analyzing the existing procedural documents and interviewing with experts, the following problems are identified:

\begin{itemize}
    \item It lacks a common and formalized vocabulary among departments which impairs data interoperability and causes ambiguity. The items with similar names may have multi-level relations, such as nested scopes of responsibilities and spacial interconnections. They may cause misunderstanding among different departments.         
    
    \item It lacks an digitized information exchange platform including documented and tacit knowledge to provide guidance for technicians and improve the work efficiency.     
    %The sampling worksheets can be  to facilitate data logging, instant processing and analysis and to support decision-making in management layer. 
    
    \item The sampling tasks placed in the daily worksheet may be not in the optimal arrangements with regards to time saving and matching with the personal working pattern of the technicians. The digital worksheet retrieved from the database of GLIMS can be reshuffled in the digital Apps to follow the most efficient sequence and the customized sampling patterns of the technicians, where GLIMS is a kind of LIMS used by E\&M department.
    
    %\item The items with similar names but belonging to different parts of the whole procedure have multi-level relations such as the nested scopes of responsibilities, spacial interconnection, and affiliations. It may confuse novice technicians during training when they manage to understand the items with similar names with regards to their functional differences and similarities. Also, as the number of the items belonging to the same topic increase, the complexity of the semantic system around this topic grows inevitably, which may cause misunderstanding due to typos in the communication and documents exchanged among different departments.   
    
    %For instance, there are several similar location items that need to be clarified for better understanding, memory and training of the technicians about their relationships and affiliations, i.e., local 4003, the SAS 4003, the SAS of local 4003 and SAS of reception 4003. And the relations between ``water network”, ``suites water”, ``water circuit” and ``UTY water” need to be clarified.
    
    \item The complete sampling procedure is divided into several segments and involves multiple functional agents (technicians, executive teams and departments). The information exchanges among these segments and agents are not formally modelled and the roles and functions of agents are not defined explicitly enough.        
    %For instances, there are no explanations on for what, by whom or in which way ``the Production is informed either by the initiator of the PE, or by the QC Support” and ``notify the QC support” above. Similarly more systematic plannings e.g., detailed time schedules, initiator-receiver pairs are required for transmission of the analytical samples. 
    
    %\item The complete story of sampling procedure has been divided into several segments with overlapped parts of slightly different descriptions, which can not provide an easy-to-follow overview of the entire procedure. 
    
    %For instance, there are overlapped parts of the description of procedures in ``Preparation of sampling material”, ``Transmission of analytical samples to QC” and ``Retrieval and storage of analytical samples” sections. The narrative mode could be improved to enhance conciseness and continuance.
    
    \item Some of the key instructions, such as the daily worksheet and sampling maps, are still paper-based. It lacks a digitized and interactive manner to support technicians performing maintenance activities.

    \item The current RACI matrix for responsibilities management, as shown in Fig.~\ref{sampling_figs_raci}, is not sufficient to provide enough details on the hierarchy and interconnections of tasks. It also needs to be digitized to enhance readability and accessibility.
    
    % \begin{itemize}
    %     \item The variants of RACI matrix could be employed to enrich the details of management to provide more details on the hierarchy and interconnections of tasks in the whole procedure and better preciseness for assignments of responsibility on a more fine-grained scale.
    %     \item Some of the tasks and events in the extant table-based RACI matrix such as ``procedures update", ``periodic review" are too general, while some of them like ``filling in the follow-up sheet for a PE 20208438" are too specific, so the responsibility assignment with appropriate granularity is needed. 
    %     \item  It can be digitized to enhance readability and accessibility, and modelled as the application-level ontologies instantiated by the responsibility assignment model and stakeholders communication structure in the ontology layer of the proposed digitalization framework above to complement the procedures and specifications of ``sampling methods" and to support the semantic-driven system integration for all the lifecycle phases.
    % \end{itemize}
\end{itemize}

Apart from the documented specifications and expert knowledge, the front-line  experience from the technicians is also collected and summarizes as a list of areas where field practice falls short:
\begin{itemize}
    \item Loss of time due to inconvenient linking between physical sampling points and paper maps, and between sampling equipment and sampling tasks. 
    %\item Lack of well-deigned spatial-temporal sequence of sampling points to follow.    
    \item Lack of efficiency for working and training due to paper-based instructions. 
    \item Minor details of sampling operations may be overlooked due to the lack of visual tools, increasing the risk of sanitation violations. 
    \item The repetitive operations may lower the attention of the technicians leading to inadvertent mistakes due to lack of alerting mechanisms.
\end{itemize}

\subsubsection{Improvement proposals}
To mitigate the above-mentioned problems, the potential improvement methods are proposed based on the digitalization framework as introduced in Section ~\ref{sec:2}.

\paragraph{(1) Digital sampling assistant}

To facilitate daily maintenance activities of the technicians and new employee onboarding, on the basis of the “Workflow Automation Module", the digital applications accessed via desktop computers, mobile phones or tablets can be developed to provide the following functions:
 \begin{itemize}
     \item Illustration of the sampling points on the digital maps: supported by the interactive UI design with diverse modes in the ``Visualization Module", the digital maps marked with all of the sampling points can be reviewed by zooming in or out. These marks are clickable and directed to an actively maintained list containing essential information, e.g., location, type of water, mechanical characteristics, and images, for each sampling point. In the page of specific sampling task, the involved sampling points will be illustrated and highlighted in the digital maps and its essential information will be showcased together.
     \item Knowledge base providing information of sampling operations: maintained by the ``Knowledge Management Module", a knowledge base containing essential information of the sampling method, e.g., equipment required, textual and video references for the key steps of sampling operations for different kinds of purified water, is actively maintained so that in the page of specific sampling task, the corresponding essential information from the knowledge base will serve as complementary key points in accordance with the involved sampling methods together with the sampling date and time. In the page of specific sampling task, there will be check-boxes for each sampling method required. 
     \item Connection to the exported worksheets from GLIMS: the established connection to data resources such as GLIMS in the ``Data Layer" can be synchronized as needed, the daily worksheet from GLIMS can be exported in specific format and imported into the digital applications. The sampling tasks involved in the worksheet will be displayed on the menu page, and after clicking, they will jump to their respective pages.
     \item Automated workflow: the worksheet data from the GLIMS can be refreshed in accordance with the selected date. Upon completion of sampling, the checked digital worksheet with other input information can be exported for further data processing and analysis in the ``Maintenance Analysis Module" to support decision making of the stakeholders.
     \item Clustering of the sampling tasks: an advanced set of functions provided by ``Decision-making\& process-control Module" to cluster the sampling tasks in the worksheet according to specific metrics such as the adjacency of the sampling points' spatial relations and the schedules of technician team.
     \item Information exchange platform: note that different from the codified knowledge written down in the procedure documents and instructions, most of the implicit knowledge can be acquired through practicing in field. So an information exchange platform can be embedded in the digital application for the technicians to leave comments and experiences during sampling activities, which will be summarized as the categorized tacit knowledge and error-prone points for the existing procedures. And these information can be further processed and analyzed in ``Maintenance Analysis Module", and support updates of ``Training Programs Module" and corresponding components in ``Knowledge Management Module".
 \end{itemize}

\paragraph{(2) Updated procedures}

The current sampling procedures can be updated regarding readability, accessibility, and digitalization in the following ways.
\begin{itemize}
    \item Build and maintain ontology based semantic model: the ontologies in the ``Maintenance Ontology Module" for maintenance activities especially for the water sampling activities conducted regularly can serve as the controlled knowledge base and a formally-structured set of vocabularies with good extensibility, supported by group collaboration. They are aimed to ensure the accumulation of data and experiences, to facilitate the annotation and integration of data resources and to enhance data interoperability with the help of ``Interoperability Module".
    \item Rewrite the operation specifications: the ``Responsibility\& communication Module" provides an easy-to-follow narrative of the entire procedure, model the communications and information exchanges between functional agents in details, enhance the demonstrations of sampling operations by means of interactive visual diagrams and workflow charts and interactive visual diagrams approaches, through which the users can be directed to corresponding animation by clicking the icons in the diagram.
    \item Good accessibility: parts of the updated procedure should have good accessibility for the technicians enabled by the user experiences parts in ``Visualization Module". For instance, the sampling operation specifications can be digitized and accessed within the digital tools so that they do not need to check the paper procedure individually.
\end{itemize}

\paragraph{(3) E-learning \& training courses}

The E-learning \& training courses provided by the ``Training Programs Module" can allow the technicians be familiar with the whole sampling procedure in an intuitive and illustrative way. They should contains the following functions.
\begin{itemize}
    \item Follow the whole sampling procedure: the training courses must cover all of the characters of involved functional agents and water sampling processes. The quiz can be designed to enhance the understanding of the functions of each process and their interconnection.
    \item Illustration of equipment: the novice technicians can not distinguish the different sampling equipment easily, especially the containers appearing similar, as shown in Fig.~\ref{sampling_figs_bottle}. Powered by the ``Visualization Module", the sampling equipment will be illustrated during E-learning and the special quiz set will be designed to enhance the ability of the technicians to distinguish them efficiently.
    \item Demonstration of steps via multimedia: the pivotal sampling operations that are error-prone during training and difficult to describe in texts can be divided into a series of steps and demonstrated in the form of multimedia such as the segmented videos, animation, immersive interactions powered by AR, VR.
\end{itemize}

\paragraph{(4) Field improvements}

The feedback and comments from the on-site operators left in the information exchange platform are retrieved and transferred to the ``Responsibility\& communication Module" for collective user requirement scheming. To facilitate sampling operators in field, the following field improvements can be made.
\begin{itemize}
    \item 3D printed carry box: optimization of the shape of carry box according to the contained containers to increase stability during technician movement and decrease the risk of misplacing the containers.
    \item MR technology support: in the future, the MR technology support can be developed to establish connection between the physical and digital worlds with the help of AR devices, e.g., the labels with QR codes attached to the on-site equipment can be recognized automatically or scanned manually by the MR technology powered mobile devices, and the information of the specific equipment can be accessed in the associated digital tools. The support platform can also display useful information for technicians when they approach a certain sampling point within certain predefined thresholds. Advanced features include assisted video and image recording, fault diagnosis and potential risk detection enabled by massively collected data, data science methods and artificial intelligence algorithms.
\end{itemize}

\subsection{Implementation}

The implementation of the proposed digitalization framework shown in Fig.~\ref{DT_RA_1} for the case study enabled by the aforementioned technologies is introduced. First, the user requirements and goals for the digital services are created, and problems with regards to GMP in water sampling activities are identified and preliminary solutions developed. Digital modules are prototyped, tested, and deployed, with databases, software, and platforms established. The modules are integrated and tested as a whole system, with feedback used for further improvements. Secondly, the modules are audited, reviewed, and revised based on operating conditions and user feedback. The application-level ontologies and knowledge bases are updated and refreshed. Finally, performance evaluation reports are used to decide whether to remove certain components of the digital framework. If modules are removed, corresponding resources are recycled. The tools used and the key functions of each digital module are further introduced as follows.

\subsubsection{Tool-chain}
%The diagram of tool-chain to support implementation of the digitalization framework for water sampling activities is shown in Fig.~\ref{tc_diagram_1}.
\begin{figure}[H]
     \hspace{-4.9mm}
  \includegraphics[width =   1.06\textwidth]{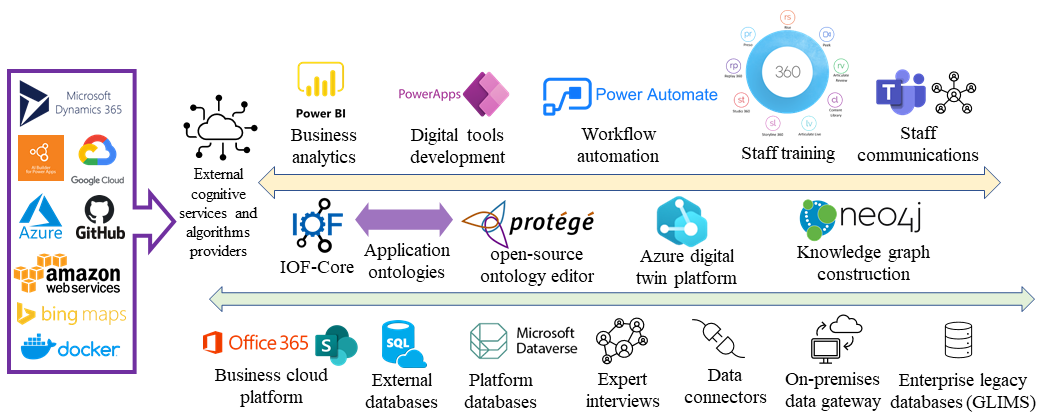}
  \caption{A tool-chain to support implementation of the digitalization framework for water sampling activities.}
  \label{tc_diagram_1}
 \end{figure}
 
The tool-chain used in this case study is illustrated in Fig.~\ref{tc_diagram_1}. It consists of a series of tools to enable the functional modules of the proposed framework which has a similar hierarchical structure as the functional layers as shown in Fig.~\ref{SDT_YZ_1}. The cloud computing platforms on the left of the tool chain, such as AWS, Azure services and Google Cloud, serves as the computing and communicating infrastructure providing cloud-based computing data storage services.  
\begin{itemize}
    \item The bottom layer contains data bases, channels, legacy platforms and software relevant to maintenance activities serving as data sources. Microsoft Office 365 provides several tools and platforms to store both scalar values and non-scalar objects.
    SharePoint is used for internal documents sharing and feedback recording. 
    GLIMS is a management system with the database storing all the laboratory information, including the daily water sampling schedules. 
    The Data Connectors serve as the bridge between different data sources and data platforms, enabling seamless data integration, transformation, and on-premises data gateway facilitates secure and efficient communication between cloud-based applications and on-premises data sources, ensuring data privacy and enabling real-time access to on-site data.

    \item The middle layer includes necessary tools, applications and frameworks to boost ontology-based knowledge management and data interoperability. An open-sourced software Protégé is used to perform application-level ontologies modelling based on IOF-Core. Neo4j can be used to build graph database and knowledge graph according to the imported ontologies.

    \item  The top layer hosts various services fulfilling the demands of different stakeholders. Power BI is used to conduct business analytics. PowerApps is used to develop web-based tools. Power Automate is used to streamline some repetitive procedures and operations. Articulate 360 serves as learning management system for staff training and provides some authoring tools. Microsoft Teams is used for internal staff communications. 
\end{itemize}
 
\subsubsection{Application-level ontologies modelling}
\begin{figure}[H]
     \hspace{.9mm}
  \includegraphics[width =   0.96\textwidth]{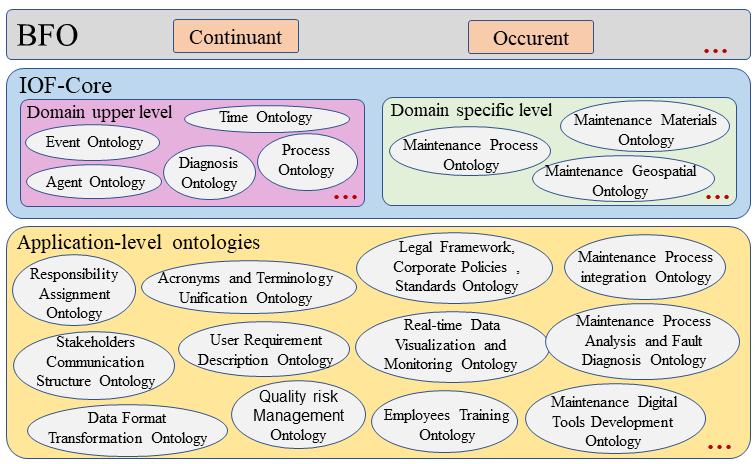}
  \caption{Overview of the ontology integration approach.}
  \label{ontology_arch_1}
 \end{figure}

The construction of the application-level ontologies refers to the ``Maintenance Ontology Module" of the ``Ontology Layer" of the digitalization framework for the water sampling activities. The overview of the ontology architecture is illustrated in Fig.~\ref{ontology_arch_1}.  

\subsubsection{Interactive E-learning Platform}
The ``Training Programs Module" contains the interactive E-learning platform. Accompanied by the authoring tools, it brings many unprecedented benefits to employee training in the era of Pharma 4.0 and they have caught the attention of leaders in maintenance activities in the pharmaceutical industry. It lifts training venue restrictions and provide highly flexible and accessible training courses. It highlights the knowledge points on safety and hygiene in every part of training courses to enable compliance of regulations, in the workplaces in the forms of multi-media demonstration. The high-level engagements (e.g., real-time statistical monitoring of learning data and assessments, tracking learning scores and progress with the leader-boards, and instant staff feedback) enhance interactivity, motivation and immersion for the staff. It can support version, format and content updates of already deployed training courses to accommodate any changes in working conditions.
\subsubsection{Digital sampling assistant}
\begin{figure}[H]
     \centering
  \includegraphics[width =\textwidth]{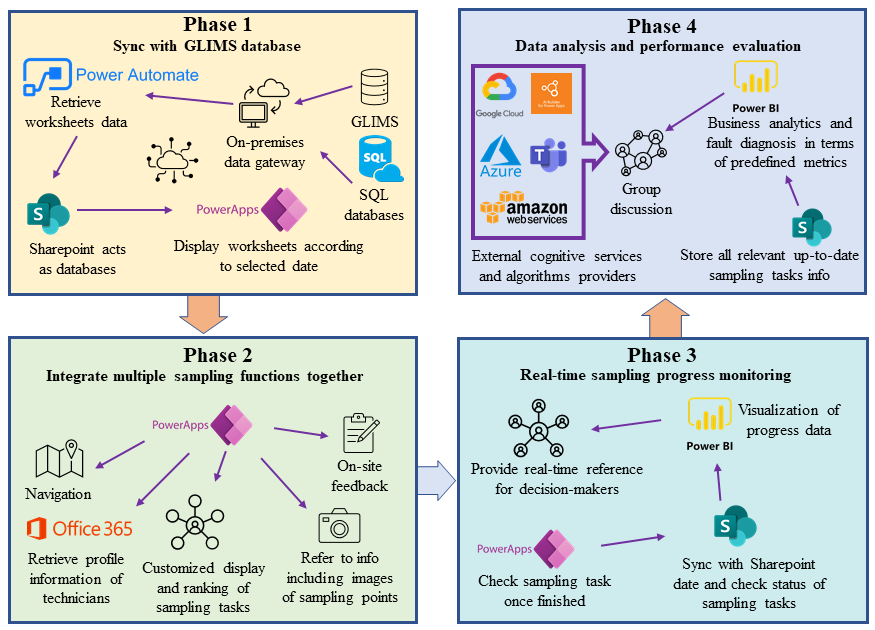}
  \caption{Workflow overview of the digital sampling assistant.}
  \label{dsa_diagram_1}
 \end{figure}

Located in the ``Workflow Automation Module" in the ``Service Layer" of the digitalization framework for water sampling activities, the digital sampling assistant integrates several processes of water sampling activities into an automated workflow. It acts as an ecosystem consisting of multiple software and platforms that provide web-based services. Its overview diagram is illustrated in Fig.~\ref{dsa_diagram_1}.

\begin{figure}[H]
\centering
\begin{subfigure}{0.475\textwidth}
    \includegraphics[width=\textwidth]{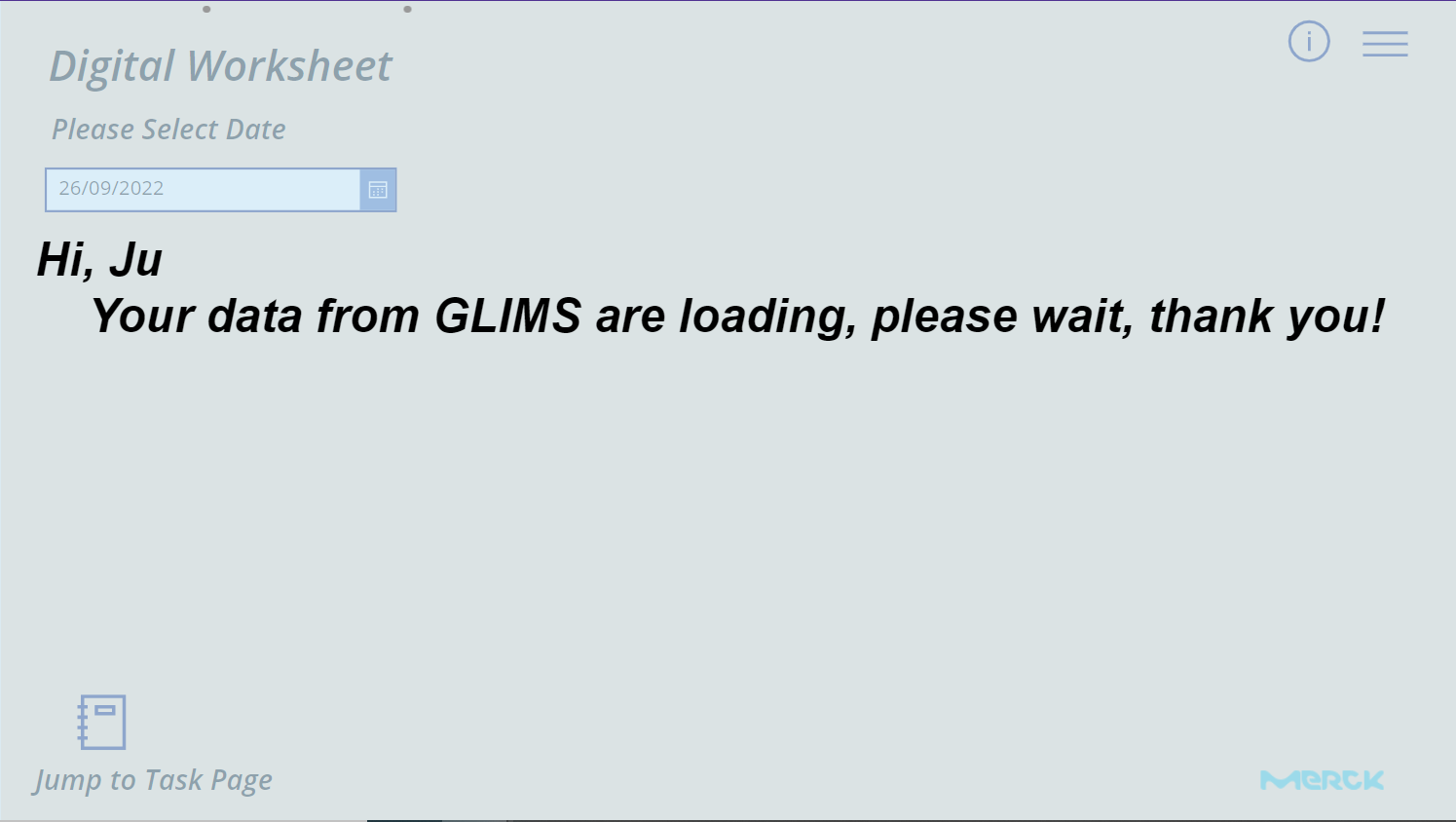}
    \caption{Welcome message at the very beginning.}
    \label{welcome_page_1}
\end{subfigure}
\hfill
\begin{subfigure}{0.475\textwidth}
    \includegraphics[width=\textwidth]{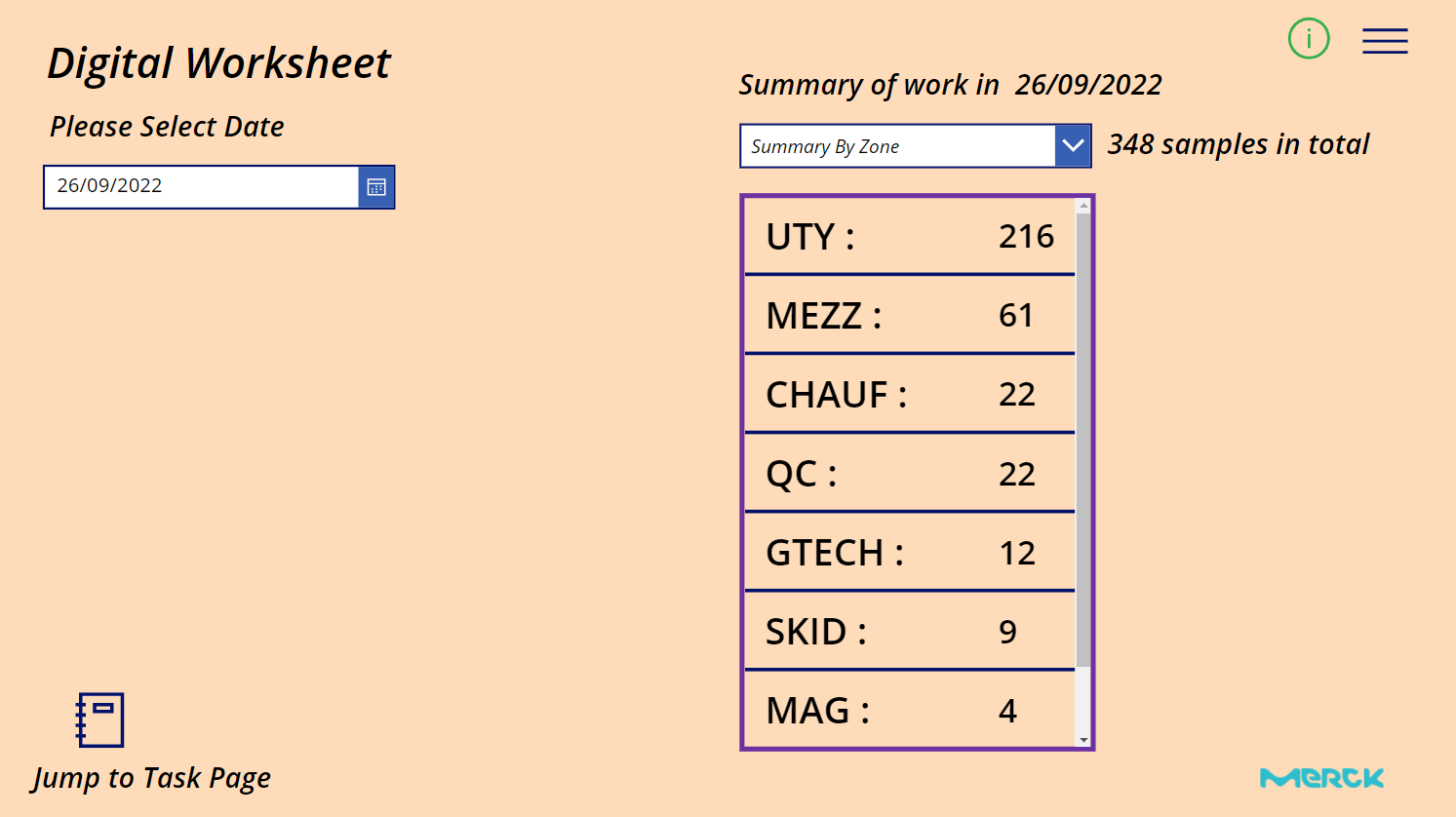}
    \caption{Customized presentation and sequencing of sampling tasks and their key information.}
    \label{menu_display_1}
\end{subfigure}
\hfill
\begin{subfigure}{0.475\textwidth}
    \includegraphics[width=\textwidth]{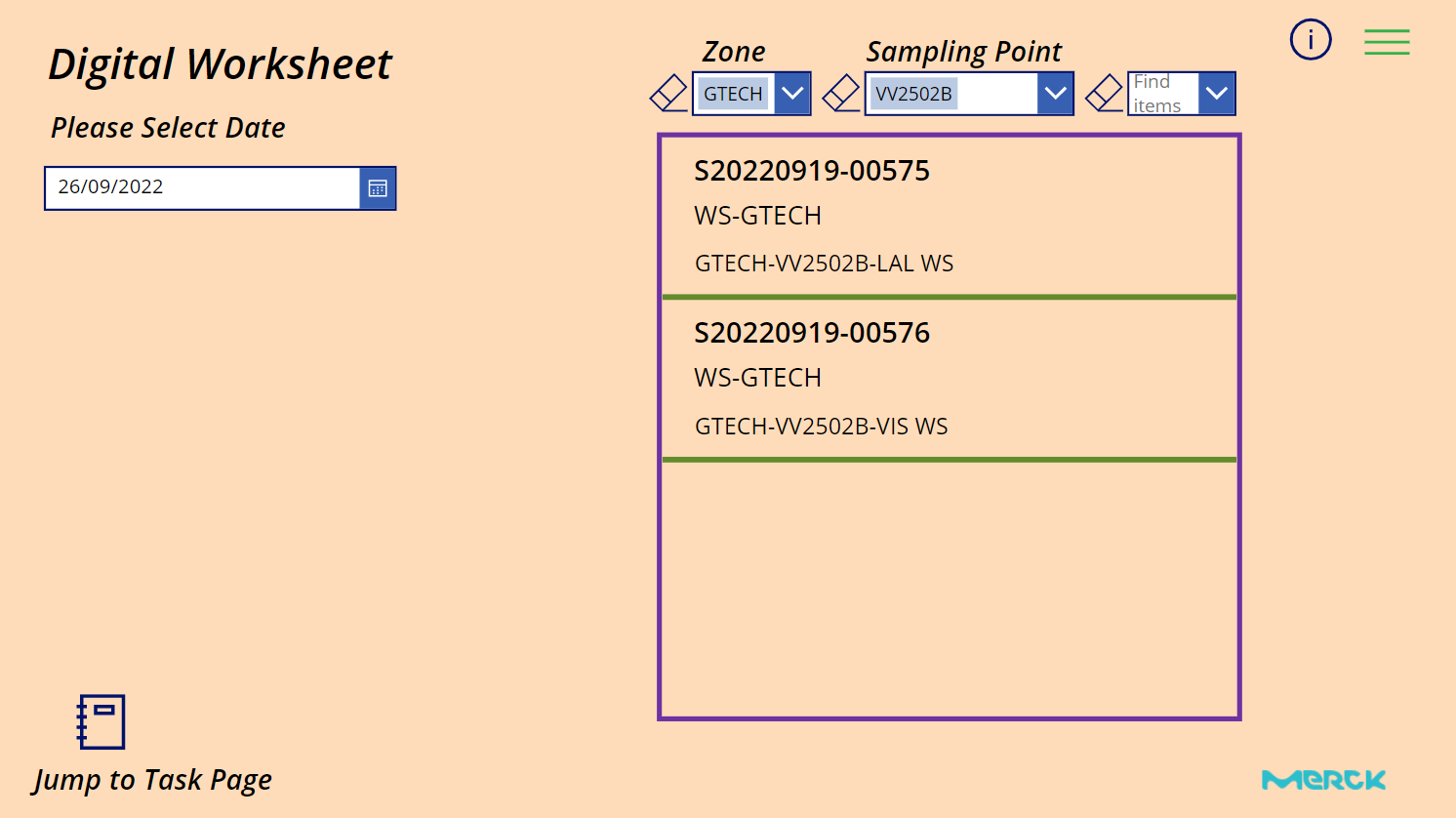}
    \caption{Custom filtering for the sampling tasks.}
    \label{menu_display_2}
\end{subfigure}
\hfill
\begin{subfigure}{0.475\textwidth}
    \includegraphics[width=\textwidth]{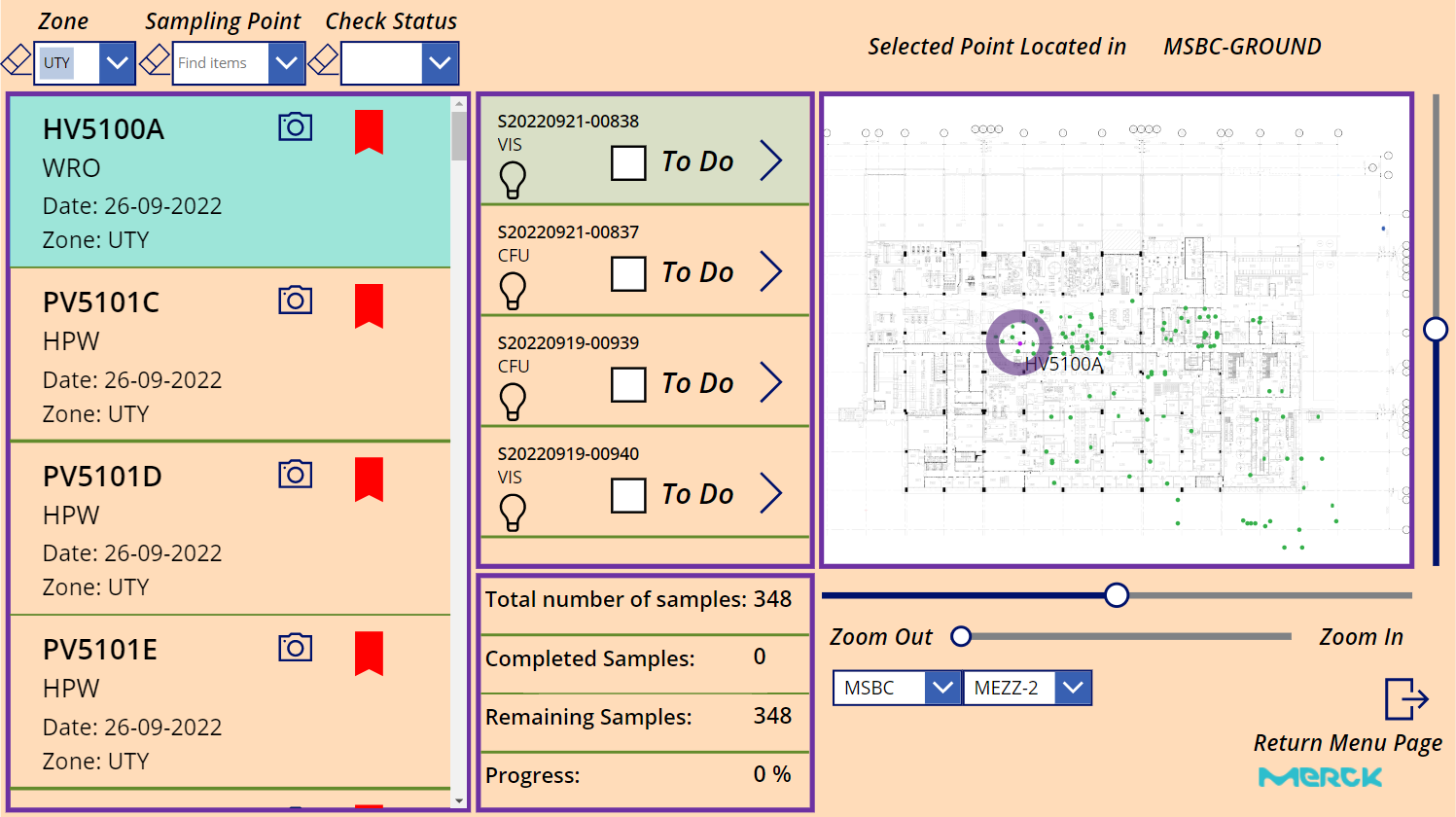}
    \caption{Sampling task page with multiple functionalities (digital maps, check-in, on-site feedback).}
    \label{task_page_1}
\end{subfigure}
\hfill
\begin{subfigure}{0.475\textwidth}
    \includegraphics[width=\textwidth]{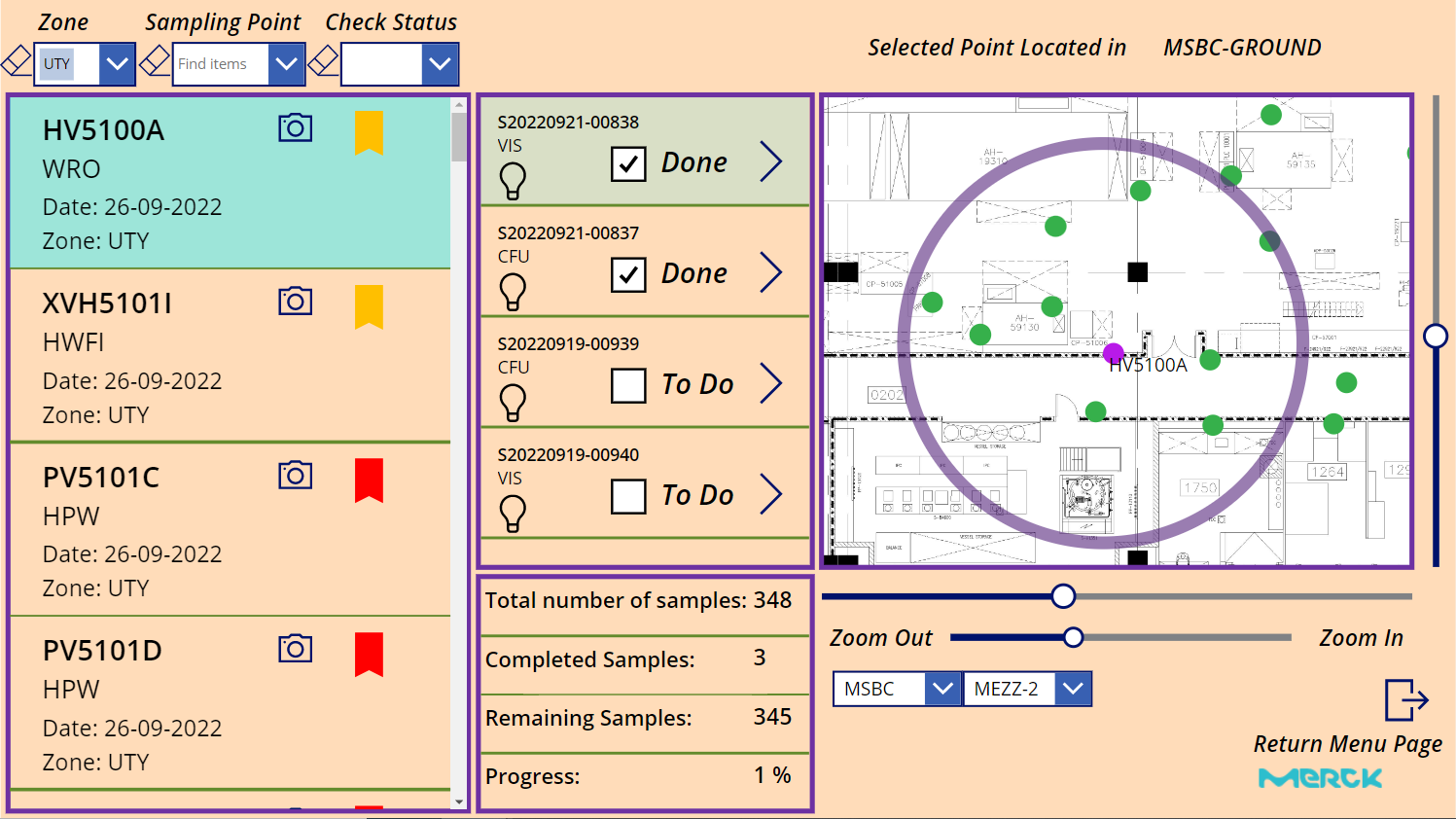}
    \caption{Zoom in and out around the center point for digital maps.}
    \label{task_page_2}
\end{subfigure}
\hfill
\begin{subfigure}{0.475\textwidth}
    \includegraphics[width=\textwidth]{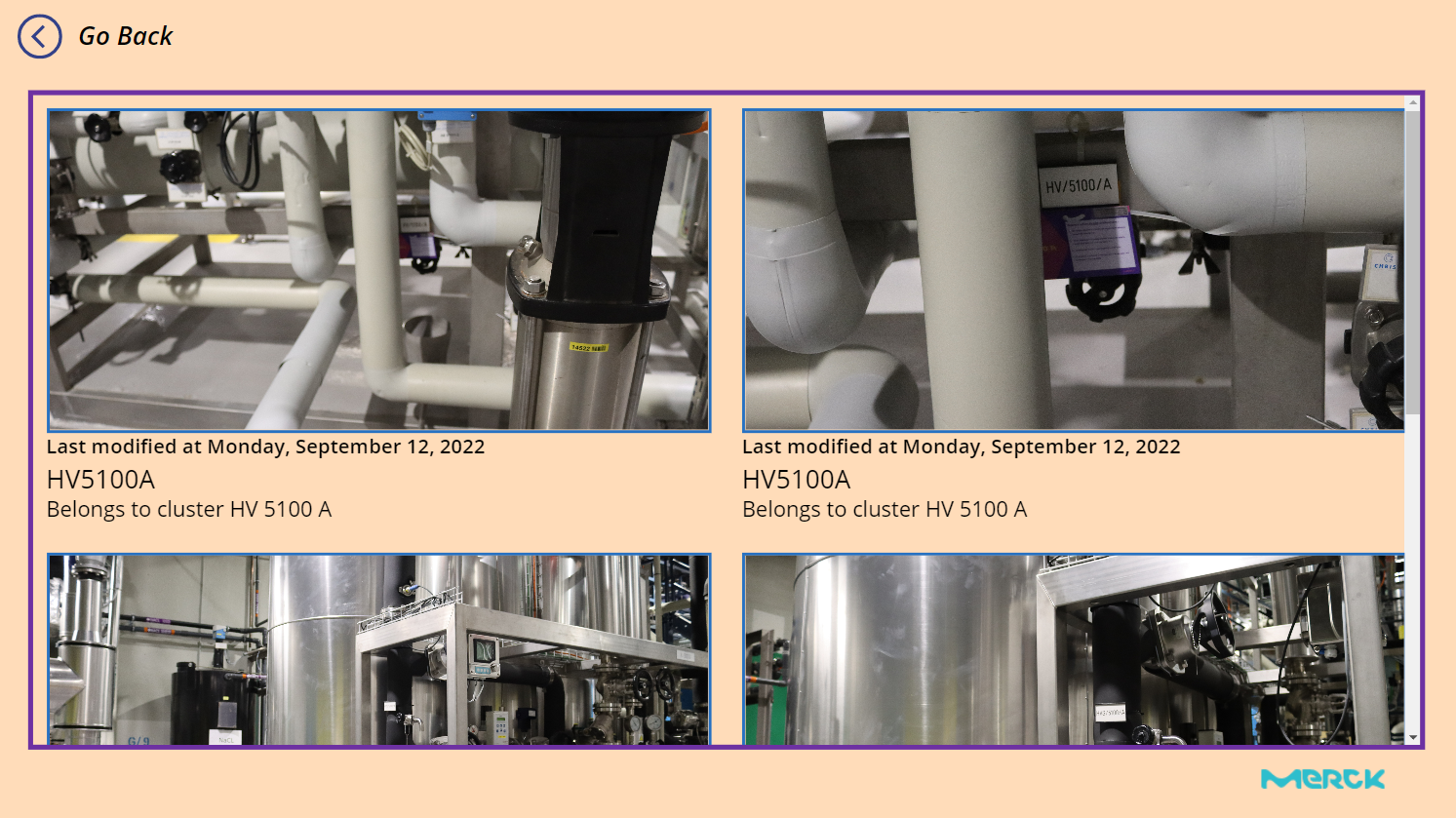}
    \caption{Useful relevant information connected to sampling points such as their scene images.}
    \label{img_info_1}
\end{subfigure}
\caption{Parts of the images of the digital sampling assistant.}
\label{dsa_imgs}
\end{figure}

In practice, when the technicians open the digital sampling assistant, the message will pop up to remind them to wait for establishment of connection with GLIMS and corresponding data synchronization, as shown in Fig.~\ref{welcome_page_1}. At default, worksheet data from the day he opens the digital sampling assistant will be retrieved, processed, and displayed. The main functionalities are briefly introduced as follows.

\begin{itemize}
    \item \textbf{Electronic worksheets}: the digital sampling assistant supports intuitive displaying and customized ranking of the sampling tasks and their associated statistics according to the column features i.e., ``Sampling Zone", ``Sampling Method", ``Sampling Point", ``Sampling Execution Date", etc, as shown in Fig.~\ref{menu_display_1}.  Specific sampling tasks can be filtered and displayed based on their column features i.e., ``Sampling Zone", ``Sampling Method", ``Sampling Point", as illustrated in Fig.~\ref{menu_display_2}.
    \item \textbf{Digital maps and scene pictures}: as shown in Fig.~\ref{task_page_1}, the interactive maps for the sampling zones are marked with the sampling points based on their geospatial information. When the technician clicks certain sampling point, it will be highlighted in the corresponding sampling zone map for easy navigation and its associated sampling tasks will be visualized in the gallery form. The technicians can search for the desired sampling points based on ``Sampling Zone", vague name of sampling point and the check-in status of sampling points (i.e., ``Untouched", ``Partial", ``Completed"). The digital maps can be zoomed in and out around the center point by changing the value of slider as shown in Fig.~\ref{task_page_2}. If a technician clicks the camera icon for a particular sample point, they will be directed to a page detailing all media related to that sample point, such as its scene pictures, as shown in Fig.~\ref{img_info_1}   
    \item \textbf{Sampling tasks check-in and on-site feedback}: real-time on-site feedback from the technicians is enabled. Once a sampling task is finished, the technician can check the corresponding checkbox to log the key information such as ``Check Status", ``Execution Time" and transfer them to the connected Sharepoint List.
    \item \textbf{Real-time progress monitoring and predictive maintenance} all of update-to-date information about the sampling tasks generated by the digital sampling assistant are centralized in the connected Sharepoint List, and Power BI establishes connection to this Sharepoint List. Apart from basic functions to monitor real-time progress of sampling tasks and evaluate sampling performance in terms of efficacy and error rate, more sophisticated algorithms for predictive maintenance such as risk assessment and fault diagnosis are developed in Power BI to support ``Decision-making \& Process-control Module", ``Visualization Module" and ``Maintenance Analysis Module" in service layer of the digitalization framework for water sampling activities.
\end{itemize}

\subsection{Discussion}

In this case study, the proposed semantic-driven digitalization framework is applied to overcome the challenges encountered during daily water sampling activities. The functional modules of the framework are implemented to satisfy the user requirements and enable improvements regarding the five pivotal GMP elements.
Digital applications supported by the tool-chain have been developed to facilitate daily maintenance activities and new employee onboarding. Functions of the digital sampling assistant in the ``Training Programs Module" include digital maps with clickable sampling points, interaction with a knowledge base for sampling operations located in the ``Knowledge Management Module", connection to exported worksheets from GLIMS of the ``Data Layer", an automated workflow powered by the algorithms in the ”Decision-making\& process-control Module” and ``Maintenance Analysis Module" for data pre/post-processing, clustering of sampling tasks, and an information exchange platform for technicians to leave comments and experiences.

The sampling procedures are updated for enhancement of readability, accessibility, and digitalization. The knowledge base has been built and actively maintained in the ``Knowledge Management Module" using the ontology-based semantic models on the basis of ``Maintenance Ontology Module” and ”Interoperability Module". The sampling operation specifications have been rewritten for easy-to-follow narrative and interactive visual aids, and digitized to boost accessibility, enabled by the ”Responsibility communication Module” and ``Visualization Module".

The training courses in the interactive E-learning platform provided by the ``Training Programs Module" have provided technicians with an intuitive and illustrative understanding of the sampling procedure. The courses cover all aspects of functional agents and water sampling processes, and include quizzes to enhance understanding. Illustrations of equipment have been provided to aid novice technicians in distinguishing between similar containers. Additionally, pivotal sampling operations that are error-prone or difficult to describe in text have been demonstrated through multimedia such as segmented videos, animations, and immersive interactions powered by AR/VR.

Based on the efficient on-site feedback from operators in information exchange platform, collective discussion and user requirements scheming for improvements supported by the ``Responsibility\& Communication Module", and analysis of the failure data in the well-maintained knowledge base, the 3D-printed carry box with optimized shape has been designed to increase stability during movement and reduce the risk of misplacing containers. The potential proposals to enhance connection between the physical and digital worlds and evolve the current integrated digital modules consistently are made by all the stakeholders and some of them have reached consensus for future development. For instance, labels with QR codes attached to on-site equipment will be scanned by mobile devices, providing technicians with information about the specific equipment; the platform will also display helpful information as technicians approach a sampling point; the assisted video and image recording, fault diagnosis, and risk detection using data science and AI algorithms will be deployed in the ``Maintenance Analysis Module".

Higher efficiency is demonstrated, for instance, technicians must carry printed or handwritten maps for reference during sampling operations; in-field events such as water or steam leakage were recorded with portable notebook to record  of the specific sampling points; and emergency accidents were reported using phone calls. The developed digital sampling assistant integrates all of the above to provide an all-in-one solution for more efficient communication and better compliance with safety and hygiene regulations; the exchange of information between the different digital modules enhances paperless operations, which benefits sustainability. 

The case study verifies the effectiveness of the proposed digitalization framework dedicated to the maintenance strategies in pharmaceutical industry and it can be applied to other specific maintenance activities. Due to limited resources and efforts, there are some limitations of this study that can be solved in the future.
\begin{itemize}
    \item During implementation of the framework, there exists the subjective resistance mindset of the staff to work with the novel digital technologies, for instance, on-site operators have become accustomed to paper files to exchange information, record daily tasks and report the emergency situations though we have the alternative digital methods.
    \item For sustainable purposes, the minimal time and cost should be taken to transform the legacy digital assets and re-organize them into the novel digitalizaton framework. The training courses should be developed to accelerate the employee's old work customs to adapt to the novel framework.
    \item Limited by the regulations of the case owner and available resources, this case study could not realize all the functional components of the proposed framework. Therefore, the case study focused on a relatively small but important task, water sampling, as example to demonstrate the key functions of the framework. Some preliminary implementation results are introduced, but more efforts are needed for large scale implementations. 
\end{itemize}

\section{Conclusion}
\label{sec:5}
This paper proposes a digitalization framework for maintenance strategies in the pharmaceutical industry aiming to solve the digitalization challenges faced in maintenance activities in era of Pharma 4.0. A semantic-driven approach is proposed to enable the functional modules hosted by three functional layers and covering the five key elements of the pharma-specific GMP, i.e. People, Process, Procedures, Premises and Equipment, and Products. A case study was conducted and the implementation results proves that the proposed approach can help improve the robustness of maintenance activities to tackle the inadvertent mistakes such as missing tasks and wrong data entry. 
Future works include extending the application scale of the case study to obtain more precise evaluation results of the proposed approach. This study mainly focused on the functional modules covering the  five GMP key elements. More work is need in the future to explore the feasibility of the proposed digitalization framework taking into consideration of  other lifecycle pahses referring to existing reference frameworks like RAMI 4.0 and TOGAF.

% \section*{Acknowledgement}
% The work presented in this paper has been partially supported by the EU H2020 project QU4LITY (825030) - Digital Reality in Zero Defect Manufacturing, and EU H2020 project FACTLOG (869951) - Energy-aware Factory Analytics for Process Industries.

% The authors would like to recognize Merck Serono colleagues for their support during the development of this work.

% \section*{Non-financial interests}
% The authors declare that they have no known competing financial interests or personal relationships that could have appeared to influence the work reported in this paper.

% \section*{Data Availability Statement }
% The authors confirm that the data supporting the findings of this study are available within the article. Additional data such as source code and software instructions are available upon request.


 \bibliographystyle{elsarticle-num} 
 %\bibliographystyle{elsarticle-harv} 
 %\bibliography{cas-refs}

\pagebreak

\appendix
\section{Figures of existing maintenance strategies}

\begin{figure}[H]
    \centering
  \includegraphics[width = 0.95\textwidth]{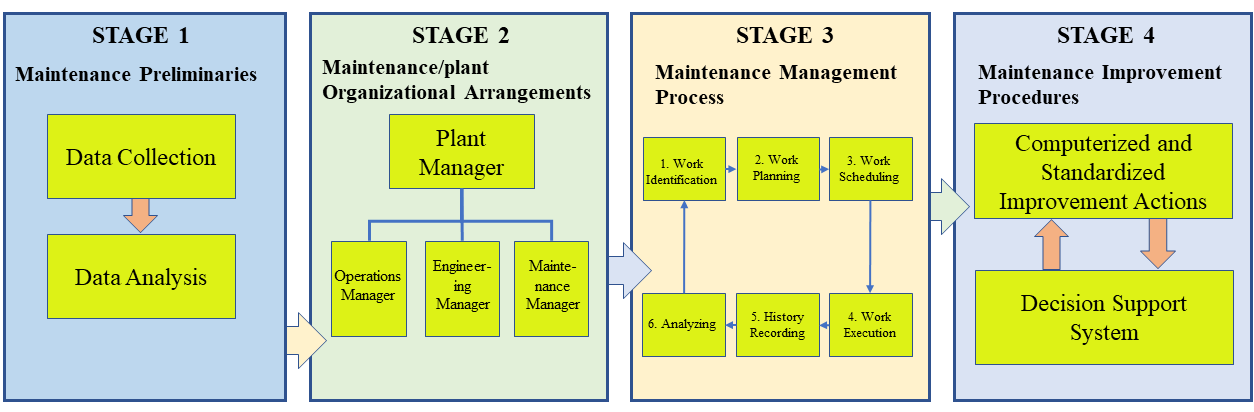}
  \caption{IMS for the maintenance strategies divided into four procedural stages \cite{felice_petrillo_autorino_2014}. }
  \label{ims_flow_1}
 \end{figure}
 
\begin{figure}[H]
    \centering
  \includegraphics[width = 0.8\textwidth]{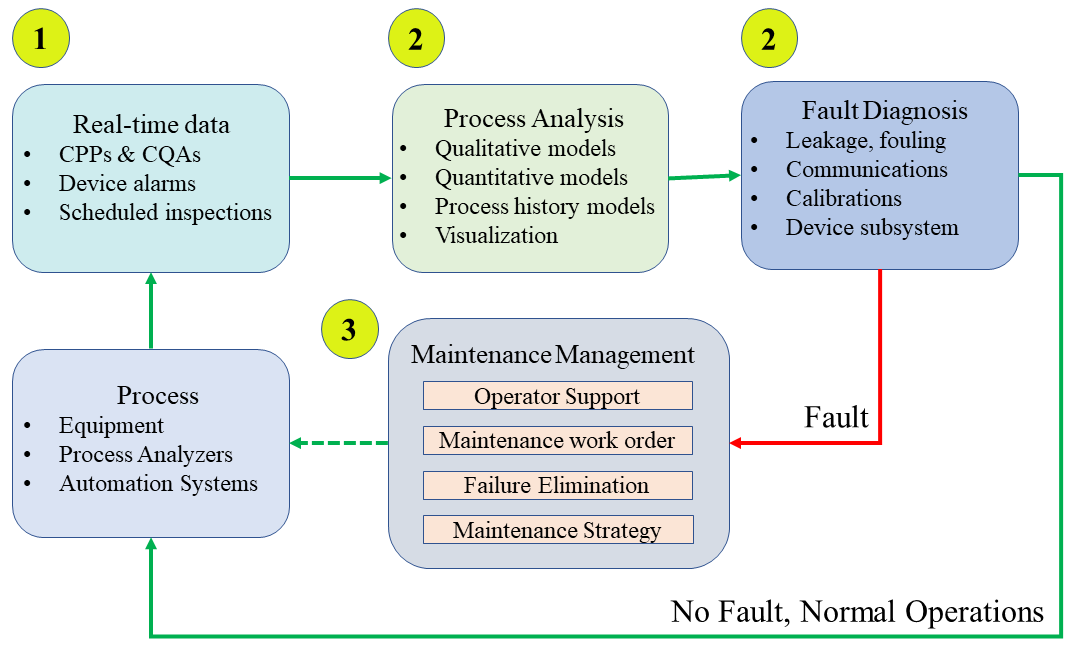}
  \caption{Data flow diagram of CBM framework \cite{ganesh2020design}.}
  \label{cbm_dataflow_1}
 \end{figure}

\begin{figure}[H]
    \centering
  \includegraphics[width = 0.5\textwidth]{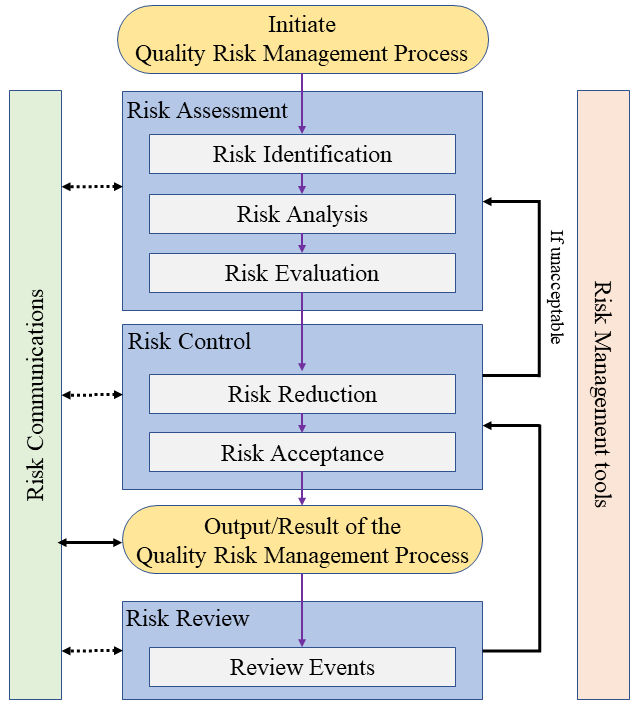}
  \caption{Overview of a typical quality risk management process \cite{guide_risk}.}
  \label{risk_mgmt_1}
 \end{figure}

\end{document}